\documentclass[12pt]{iopart}
\usepackage{xcolor}
\usepackage{graphicx}
\usepackage{dcolumn}
\usepackage{bm}
\usepackage{multirow}
\usepackage[numbers,sort&compress,square]{natbib}
\usepackage{hyperref}
\usepackage{subcaption}
\usepackage{verbatim}
\usepackage{iopams} 
\usepackage{acronym}
\usepackage[mathlines,running]{lineno}
\usepackage{tabularx}
\usepackage{listings}

\definecolor{chmagenta}{rgb}{0.54, 0.17, 0.88}

\acrodef{LIGO}[LIGO]{Laser Interferometric Gravitational-Wave Observatory}
\acrodefplural{LIGO}[LIGO]{Laser Interferometric Gravitational-Wave Observatories}
\acrodef{LVK}[LVK]{LIGO Scientific, Virgo and KAGRA}
\acrodef{SNR}[SNR]{signal-to-noise ratio}
\acrodef{CNN}[CNN]{convolutional neutral network}
\acrodef{CSV}[CSV]{comma-separated value}
\acrodef{GPS}[GPS]{Global Positioning System}
\acrodef{GWOSC}[GWOSC]{Gravitational Wave Open Science Center}
\acrodef{ReLU}[ReLU]{rectified linear unit}

\newcommand{\LSU}{Department of Physics, Louisiana State University, 202 Nicholson Hall
Baton Rouge, LA 70803 USA}
\newcommand{\CIERA}{Center for Interdisciplinary Exploration and Research in Astrophysics (CIERA), Department of Physics and Astronomy, Northwestern University, 1800 Sherman Ave, Evanston, IL 60201, USA}
\newcommand{\NUIT}{Northwestern University Information Technology Research Computing Services, Northwestern University, 1800 Sherman Ave, Evanston, IL 60201, USA}
\newcommand{\MIT}{LIGO, Massachusetts Institute of Technology, Cambridge, MA 02139, USA}
\newcommand{\KICP}{Kavli Institute for Cosmological Physics, The University of Chicago, 5640 South Ellis Avenue, Chicago, IL 60637, USA}
\newcommand{\EFI}{Enrico Fermi Institute, The University of Chicago, 933 East 56th Street, Chicago, IL 60637, USA}
\newcommand{\SUPA}{SUPA, School of Physics and Astronomy, University of Glasgow, Kelvin Building, University Ave, Glasgow G12 8QQ, UK}
\newcommand{\Syracuse}{School of Information Studies, Syracuse University, 343 Hinds Hall, Syracuse, NY 13210, USA}

\newcommand{\CSUF}{Nicholas and Lee Begovich Center for Gravitational-Wave Physics and Astronomy (GWPAC), Department of Physics, California State University Fullerton, Fullerton, 800 North State College Blvd, CA 92831, USA}
\newcommand{\EECS}{Electrical and Computer Engineering, Northwestern University, 2145 Sheridan Road, Evanston, IL 60208, USA}
\newcommand{\Adler}{Zooniverse, The Adler Planetarium, 1300 South DuSable Lake Shore Drive, Chicago, IL, 60605, USA}
\newcommand{\MS}{Microsoft Corporation, Mountain View, CA, USA}
\newcommand{\MSS}{Microsoft Corporation, Redmond, WA, USA}

\begin{document}

\title[Gravity Spy O3 data set]{Data quality up to the third observing run of Advanced LIGO: Gravity Spy glitch classifications}
\author{
J~Glanzer$^{1}$, 
S~Banagiri$^{2}$, 
S~B~Coughlin$^{2,3}$, 
S~Soni$^{4}$, 
M~Zevin$^{5,6}$, 
C~P~L~Berry$^{7,2}$,
O~Patane$^{8}$,
S~Bahaadini$^{9}$,
N~Rohani$^{10}$,
K~Crowston$^{11}$,
V~Kalogera$^{2}$,
C~Østerlund$^{11}$,
L~Trouille$^{12}$ 
and 
A~Katsaggelos$^{13,2}$
}
\address{$^1$\LSU}
\address{$^2$\CIERA}
\address{$^3$\NUIT}
\address{$^4$\MIT}
\address{$^5$\KICP}
\address{$^6$\EFI}
\address{$^7$\SUPA}
\address{$^8$\CSUF}
\address{$^9$\MS}
\address{$^{10}$\MSS}
\address{$^{11}$\Syracuse}
\address{$^{12}$\Adler}
\address{$^{13}$\EECS}

\ead{christopher.berry.2@glasgow.ac.uk}
\vspace{10pt}
\begin{indented}
\item[]
\end{indented}

\begin{abstract}
Understanding the noise in gravitational-wave detectors is central to detecting and interpreting gravitational-wave signals. 
Glitches are transient, non-Gaussian noise features that can have a range of environmental and instrumental origins. 
The Gravity Spy project uses a machine-learning algorithm to classify glitches based upon their time--frequency morphology. 
The resulting set of classified glitches can be used as input to detector-characterisation investigations of how to mitigate glitches, or data-analysis studies of how to ameliorate the impact of glitches. 
Here we present the results of the Gravity Spy analysis of data up to the end of the third observing run of Advanced LIGO. 
We classify {$233981$} glitches from LIGO Hanford and {$379805$} glitches from LIGO Livingston into morphological classes. 
We find that the distribution of glitches differs between the two LIGO sites. 
This highlights the potential need for studies of data quality to be individually tailored to each gravitational-wave observatory.
\end{abstract}

%
%
\submitto{\CQG}

\section{Introduction}\label{sec:intro}

Gravitational-wave astronomy provides unique information about our Universe. 
To date, the Advanced \ac{LIGO}~\cite{TheLIGOScientific:2014jea} and Advanced Virgo~\cite{TheVirgo:2014hva} detectors have observed signals from coalescing binaries of neutron stars and black holes~\cite{Abbott:2016blz,LIGOScientific:2018mvr,Abbott:2020niy,LIGOScientific:2021usb,LIGOScientific:2021djp}, with the rate of discovery increasing dramatically as the sensitivity of the detector network improves. 
Analysis by the \ac{LVK} Collaboration identified $3$ candidates with a probability of astrophysical origin greater than $50\%$ in the first observing run (O1) of the advanced-detector network~\cite{LIGOScientific:2016dsl}, $8$ in the second observing run (O2)~\cite{LIGOScientific:2018mvr}, and $79$ in the third observing run (O3)~\cite{LIGOScientific:2021usb,LIGOScientific:2021djp}. 
Such observations require measurements equivalent to fractional changes in distance of  $\lesssim10^{-21}$~\cite{Thorne:1987af}, and hence the detector must be carefully isolated from instrumental and environmental sources of noise. 
However, noise cannot be fully eliminated, and to identify and analyse gravitational-wave signals it is necessary to understand the properties of noise in the gravitational-wave detectors~\cite{LIGOScientific:2019hgc}. 

Transient, non-Gaussian bursts of noise (typically less than a few seconds in duration) in the gravitational-wave data stream are known as \emph{glitches}. 
Glitches are particularly detrimental to the identification and analysis of gravitational-wave signals~\cite{Canton:2013joa,TheLIGOScientific:2017lwt,Pankow:2018qpo,Powell:2018csz,LIGOScientific:2019hgc,Chatziioannou:2021ezd,Payne:2022spz}. 
There are many different glitch types, some with known environmental or instrumental origins, and others with uncertain or unknown sources~\cite{TheLIGOScientific:2016zmo,Nuttall:2018xhi,Cabero:2019orq,Davis:2021ecd,Davis:2022dnd}. 
Identifying the causes of glitches is key to improving gravitational-wave data quality.

A wide range of tools are used to monitor data quality and characterise the behaviour of the detectors~\cite{Davis:2021ecd,KAGRA:2020agh,Acernese:2022jes,hveto,gwdetchar,Robinet:2015om,Robinet:2020lbf,Davis:2022dnd}. 
In recent years, machine-learning methods have been developed for a range of analyses connected to various aspects of detector characterisation~\cite[e.g.,][]{Cuoco:2020ogp,Biswas:2013wfa,Tiwari:2015ofa,Mukund:2016thr,Cavaglia:2018xjq,Razzano:2018fxb,Vajente:2019ycy,Biswas:2019wmx,Colgan:2019lyo,Essick:2020qpo,Ormiston:2020ele}. 
The Gravity Spy project~\cite{Zevin:2016qwy,Bahaadini:2018git,Coughlin:2019ref,Soni:2021cjy} aims to classify glitches by combining human and machine-learning classification schemes: volunteers on the Zooniverse citizen-science platform (as well as \ac{LVK} detector-characterisation experts) inspect and classify individual glitches, which can then be used as input to a machine-learning algorithm that can classify large sets of data.%
\footnote{Gravity Spy Zooniverse project \href{http://gravityspy.org/}{gravityspy.org}.}  
Since its launch in October 2016, the Gravity Spy project has analysed almost {2 million} individual glitches and has accumulated over {5.7 million} classifications by more than {27,000} registered Zooniverse users.%
\footnote{The European Gravitational Observatory run a similar project dedicated to understanding glitches in Virgo data: GWitchHunters~\cite{DiRenzo:2022cxa} \href{https://www.zooniverse.org/projects/reinforce/gwitchhunters}{www.zooniverse.org/projects/reinforce/gwitchhunters}.}  
Results of machine-learning and volunteer classifications have been made available both internally within the \ac{LVK}, and to the wider public~\cite{bahaadini_sara_2018_1476156,coughlin_scott_2018_1476551,coughlin_scott_2021_5649212,michael_zevin_2022_5911227}.

Compiling a catalogue of classified glitches is useful for both identifying the physical causes of glitches (such that commissioning work could be done to remove them), and evaluating the impact of glitches on data analysis (creating new analyses to mitigate their effect where necessary). 
For example, Gravity Spy classifications have been used for: 
selecting example glitches to evaluate their impact on data analysis~\cite{Davis:2020nyf,Ashton:2021tvz,Macas:2022afm,Hourihane:2022doe}; studying glitch morphology~\cite{Torres-Forne:2020eax,Merritt:2021xwh,Lopez:2022lkd,Powell:2022pcg}; cross-referencing glitches with environmental-noise or auxiliary-channel measurements~\cite{Soni:2020rbu,Davis:2021ecd,Longo:2021avq,Colgan:2022vdd}, and as a component of training for gravitational-wave detection algorithms~\cite{Benko:2020syv,Marianer:2020slp,Jadhav:2020oyt,Singh:2020yau,Abbott:2021cuf,Chaturvedi:2022suc,Choudhary:2022yje} or glitch-classification algorithms~\cite{George:2017fbn,Cavaglia:2018xjq,Sankarapandian:2021qun,Sakai:2021rks,Yan:2022spw}. 
Additionally, identification of new classes can indicate new sources of noise and suggest areas for further commissioning~\cite{Soni:2021cjy}.

In this paper we describe the glitch classifications from Gravity Spy's machine-learning analysis of data from the first three observing runs of Advanced \ac{LIGO}; this analysis uses the Gravity Spy \ac{CNN} models previously developed for O1--O2~\cite{Zevin:2016qwy,Bahaadini:2018git} and O3~\cite{Soni:2021cjy}. 
In Section~\ref{sec:methods} we describe the gravitational-wave strain data, the machine-learning algorithm and the glitch classes; further details of the different classes used for analysis of each observing run are given in \ref{ap:classes}.
In Section~\ref{sec:results} we illustrate how results of classifications from across the observing runs can be used for detector characterisation, summarising the rates of different glitches, and highlighting results from times near potential gravitational-wave candidates; we also give an overview of the data release. 
In Section~\ref{sec:discussion} we review the implications of our results, before summarising in Section~\ref{sec:summary}. 
The data release is available from Zenodo~\cite{coughlin_scott_2021_5649212}, and the volunteer classifications~\cite{michael_zevin_2022_5911227} will be discussed in a companion paper.

\section{Methods}\label{sec:methods}

\subsection{Detector data \& detector characterisation}\label{sec:strain}

The two \ac{LIGO} detectors in the USA (Hanford and Livingston)~\cite{TheLIGOScientific:2014jea}, the Virgo detector in Italy~\cite{TheVirgo:2014hva} and the KAGRA detector in Japan~\cite{KAGRA:2018plz}, are highly sensitive instruments designed and operated for the direct detection of gravitational waves. 
The primary data output of these observatories is the strain measured by the interferometers~\cite{LIGOScientific:2019lzm}, which will contain gravitational-wave signals as well as various sources of noise; however, there are additionally many auxiliary channels of data that record the internal state of the detectors and monitor their environments~\cite{TheLIGOScientific:2016zmo,Nguyen:2021ybi,Acernese:2022ozw}.
Since the beginning of O1 in September 2015, three observing runs have been completed~\cite{Abbott:2020qfu}. 
These are preceded and interleaved with engineering runs that are used to test the performance of the detectors, and potentially diagnose data-quality issues. 
Each successive observing run is characterised by detector improvements that lead to higher sensitivity~\cite{LIGOScientific:2016emj,Abbott:2016xvh,LIGOScientific:2017bnn,Buikema:2020dlj} and, consequently, more detections~\cite{LIGOScientific:2021djp}, as well as revealing new sources of noise. 

The data quality of these ground-based gravitational-wave detectors is impacted by multiple sources of noise. 
Broadly, noise in the detectors consists of stationary Gaussian noise sources (which include quantum noise, seismic noise and thermal noise), and non-Gaussian noise sources~\cite{LIGOScientific:2019hgc,Abbott:2016xvh,Nguyen:2021ybi,Buikema:2020dlj}. 
Non-Gaussian noise includes long-lived spectral lines~\cite{LSC:2018vzm} and shorter-duration transient glitches~\cite{Davis:2021ecd,KAGRA:2020agh,Acernese:2022jes}. 
Monitoring the status of data quality, identification and mitigation of transient noise are some of the tasks referred to as detector characterisation~\cite{TheLIGOScientific:2016zmo,Davis:2022dnd}. 
Understanding and improving data quality is central to extracting astrophysical information from detector data.

\begin{figure}
    \centering
    \includegraphics[width=0.85\textwidth]{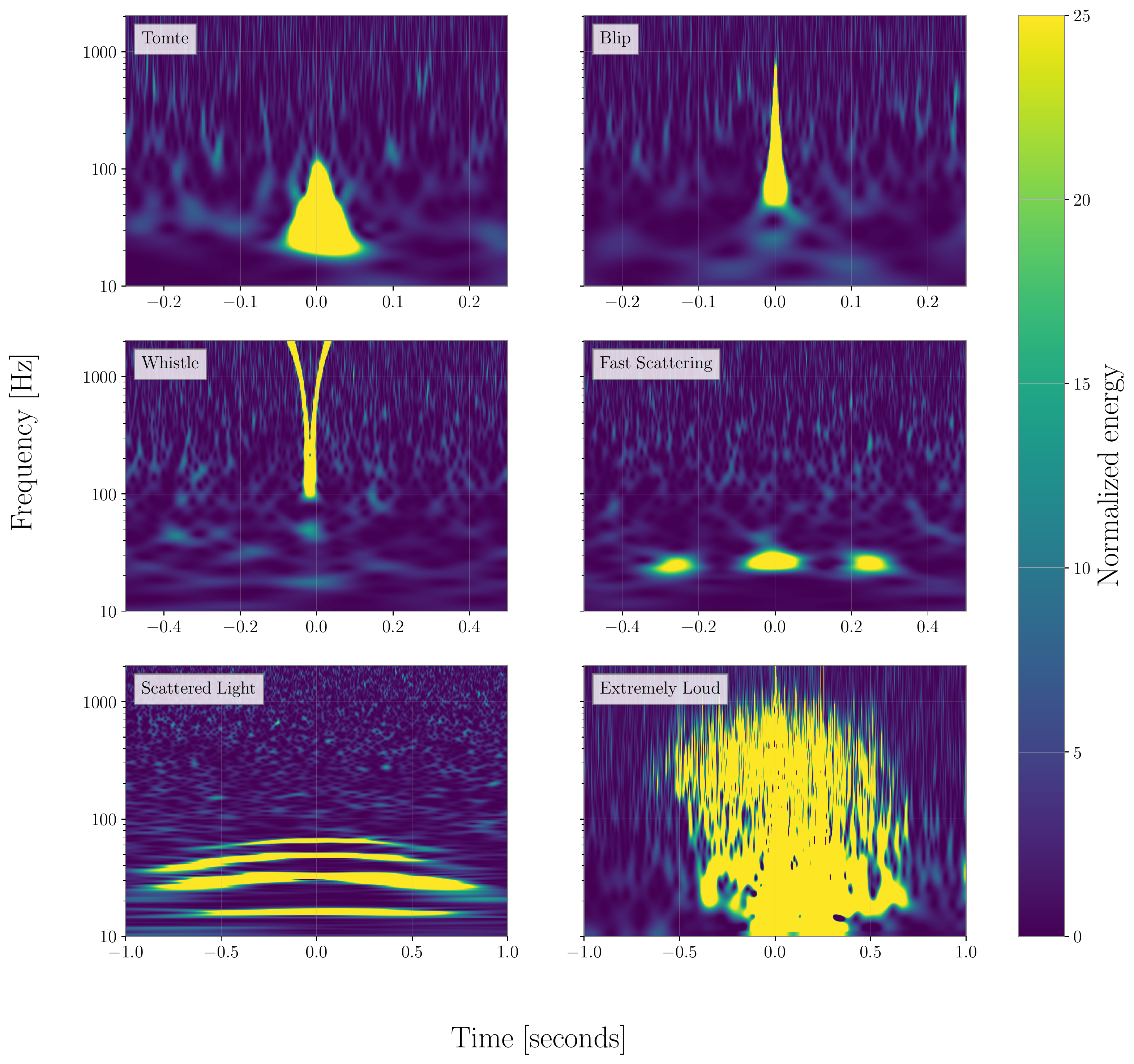}
    \caption{
    Example time--frequency spectrograms~\cite{Chatterji:2004qg} for a selection of \ac{LIGO} glitch classes. 
    The glitch classes here are relatively common and illustrate the range of morphologies different glitch classes can have. 
    The spectrograms in each row are shown with a different time duration.
    \emph{Top left:} Tomte is a short-duration glitch with a characteristic triangular morphology. 
    \emph{Top right:} Blip is another short-duration glitch, but covers a broader frequency range than Tomte and has a tear-drop morphology.
    \emph{Middle left:} Whistles have a characteristic V, U or W shape sweeping through higher frequencies ($\gtrsim128~\mathrm{Hz}$). 
    \emph{Middle right:} Fast Scattering (also known as Crown) appears as one or more arches, each $\sim0.2$--$0.3~\mathrm{s}$ in duration.
    \emph{Bottom left:} Scattered Light (also known as Slow Scattering) appears as longer-duration ($\sim2.0$--$2.5~\mathrm{s}$) arches, with multiple arches often being stacked on top of each other.
    \emph{Bottom right:} Extremely Loud are high-\ac{SNR} triggers that saturate the spectrogram. 
    Exemplar spectrograms for each Gravity Spy class are given in Figure~\ref{fig:allglitch}.
    }
    \label{fig:glitches_combined}
\end{figure}

Potential glitches (as well as gravitational-wave signals) are identified by searching for excess power in the data stream. 
All the noise transients analyzed in this paper were detected by the Omicron algorithm~\cite{Robinet:2015om,Robinet:2020lbf} analysing the gravitational-wave strain channel (and not using auxiliary channels). 
Omicron identifies potential noise transients by triggering on excess power in the data stream. 
The Omicron algorithm annotates each identified transient with characteristics such as event time, peak frequency, central frequency and \ac{SNR}. 
The glitch morphology of the trigger can be visualized in a time--frequency spectrogram commonly known as an Omega scan~\cite{Chatterji:2004qg,gwdetchar}.
These Omega scans are used frequently in data-quality studies to establish potential noise correlations between different parts of the detector~\cite{alog:transmon}, and the time--frequency morphology can be used to categorise glitches~\cite{Bahaadini:2018git,Davis:2021ecd}. 
The morphology may contain clues to the cause of the glitch~\cite{Davis:2022dnd}, e.g., arches are characteristic of light scattering, with the frequency encoding information about the relative motion of the scattering source, and multiple stacked arches suggesting repeated reflections of stray light from the scattering source~\cite{Accadia:2010zzb,Valdes:2017xce,Soni:2020rbu}.
Example Omega scans for common glitch classes are shown in Figure~\ref{fig:glitches_combined}.
These time--frequency spectrograms are used as the input to Gravity Spy.

\subsection{Machine-learning algorithm \& glitch classes}\label{sec:ML}

Gravity Spy contributes to detector characterisation by classifying glitches. 
The morphological classes used in Gravity Spy for \ac{LIGO} data are detailed in \ref{ap:classes}. 
Classifications are made based upon time--frequency spectrograms, using two complementary approaches: visual inspection by Zooniverse volunteers, and automated analysis by a machine-learning algorithm~\cite{Zevin:2016qwy,Coughlin:2019ref,Soni:2021cjy}. 
Both approaches use the same input: Omega scans of four different temporal resolutions ($0.5~\mathrm{s}$, $1~\mathrm{s}$, $2~\mathrm{s}$ and $4~\mathrm{s}$ in duration, centred on the time of the transient). 
Here we concentrate on the machine-learning classification as opposed to volunteer classification.

Gravity Spy uses a \ac{CNN}, a deep-learning algorithm used primarily for image classification, to analyse the Omega scans. 
For every image input to the \ac{CNN}, the probability (or \emph{confidence}) $p$ of belonging to each class is calculated, and the glitch is assigned to the class with the highest associated confidence~\cite{Zevin:2016qwy}.
\ac{CNN} architectures include an input layer, an output layer, and various hidden layers in between that transform the data and extract useful features. 
The \ac{CNN} used by Gravity Spy~\cite{Bahaadini:2017dqg} has four convolutional layers to extract features, each followed by a max-pooling and a \ac{ReLU} activation layer, and then a final fully connected layer and a softmax layer. 
The weights from the last softmax layer are the confidence scores for each of the classes. 
Confidence scores for each trigger, indicating the probability that it is associated with various morphological classes, are provided in the data release. 
The accuracy of the classification is tested during training of the \ac{CNN}~\cite{Bahaadini:2017dqg,Zevin:2016qwy,Soni:2021cjy}.

\subsection{The training sets}

The original \ac{LIGO} data set used to train the Gravity Spy \ac{CNN} was created by detector-characterisation experts and Gravity Spy volunteers. 
It initially contained $7718$ glitch samples from $20$ classes prevalent in the detector during O1 and the preceding engineering runs~\cite{Zevin:2016qwy}. 
These classes included No Glitch, for when no significant excess power is visible in the Gravity Spy spectrograms, and None of the Above, which was intended to catch glitches that did not fit into the other classes.
The training set was refined and updated to include the 1080 Lines and 1400 Ripples classes, which were identified by volunteers~\cite{Bahaadini:2018git}. 
This gave a training set that included {$7932$} glitch samples from $22$ classes~\cite{coughlin_scott_2018_1476551}. 
The resulting training accuracy was $98.2\%$~\cite{Bahaadini:2018git}. 
This \ac{CNN} model has been used to classify data from O1 and O2.

During O3, the presence of two new prevalent glitch morphologies motivated the addition of the Fast Scattering (also known as Crown) and Blip Low Frequency (also known as Low-frequency Blip) classes to the machine-learning model;  
in addition, the None of the Above class was removed for the final analysis, as it was decided that it was more effective for the \ac{CNN} to label such triggers with low confidence than to try to construct a class of many morphologically diverse glitches~\cite{Soni:2021cjy}.%
\footnote{None of the Above remains an option for Zooniverse volunteers. 
We anticipate that reinstating the None of the Above class may be useful for identifying new classes in preliminary analysis of future observing runs. 
Prior to the introduction of the Fast Scattering class, there were a large number of None of the Above classifications for O3 data with the characteristic Fast Scattering morphology~\cite{Soni:2021cjy}.} 
Adding in the new classes, and more examples from existing classes, this current training data set contains {$9631$} glitch samples distributed over $23$ classes, of these {$8427$} were used for training and {$1203$} were used for validation. 
The resulting training and validation accuracies were $99.9\%$ and $98.8\%$, respectively~\cite{Soni:2021cjy}. 
This \ac{CNN} model has been used to classify data from O3.


The performance of the \ac{CNN} model depends upon the quantity and quality of examples from each glitch class in the training set. 
Augmenting the training set with additional glitches classified by volunteers~\cite{michael_zevin_2022_5911227} is expected to improve the results of future \ac{CNN} models.

\section{Results}\label{sec:results}

The Gravity Spy glitch classifications can be used as inputs for a range of analyses, and here we illustrate their use as the base for detector-characterisation studies concentrating on O3. 
In Sec.~\ref{sec:classifications} we show how the distribution of glitches may be studied, and in Sec.~\ref{sec:candidates} we illustrate how data quality at specific times may be studied using the example of times around gravitational-wave candidates. 
For use in further studies, the release of the Gravity Spy machine-learning classification data set is described in Sec.~\ref{sec:data-release}.

\subsection{Glitch classifications}\label{sec:classifications}

\begin{figure}
    \centering
    \includegraphics[width=6.2in]{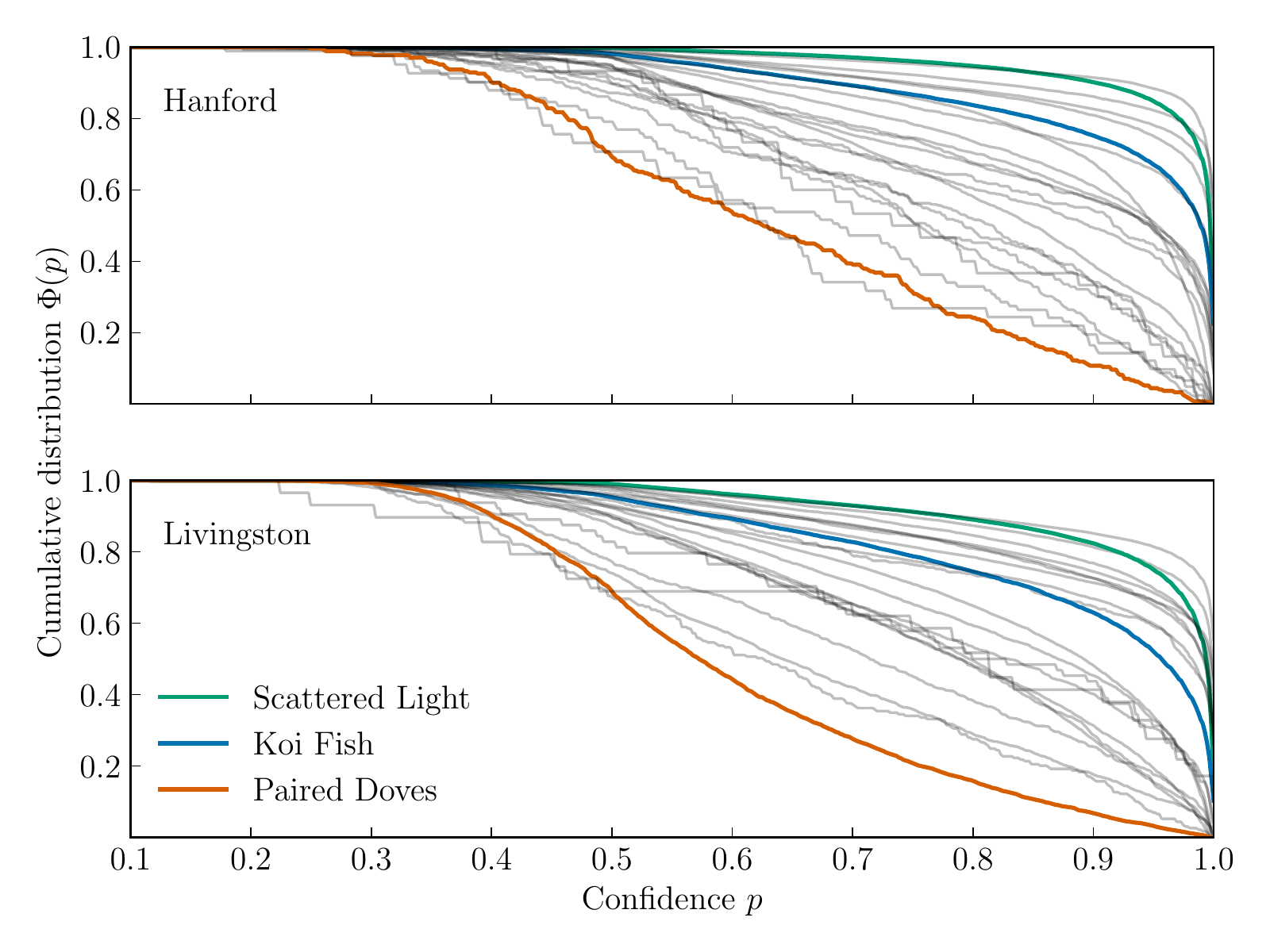}
    \caption{The cumulative distribution of O3 triggers across all classes as a function of classification confidence. 
    The horizontal axis is the confidence $p$, while the vertical axis $\Phi(p)$ is the fraction of glitches identified with confidence \emph{greater} than $p$. 
    Three glitch classes are highlighted as examples: Paired Doves (an uncommon class, with few training examples~\cite{Zevin:2016qwy,Bahaadini:2018git}), Koi Fish (a more common class, which can be confused with Blips when quiet, and Extremely Loud when loud~\cite{Bahaadini:2018git,Soni:2021cjy}), and Scattered Light (one of the most common glitch types for both detectors~\cite{Soni:2021cjy}). 
    The number of triggers in each class with $p >0.9$ and $p > 0.95$ are quoted in Table~\ref{tab:O3_glitches}.}
    \label{fig:CDF_plot}
\end{figure}

For data from both \ac{LIGO} detectors, we find that there are certain glitch classes that are more common than others. 
For example, Table~\ref{tab:O3_glitches} provides numbers of glitches sorted into the various classes from O3 data. 
In addition to the number of glitches in each class with an SNR $>7.5$, we also show those classified with a confidence $>90\%$ and $>95\%$. 
Using a higher confidence level gives a higher purity, but smaller sample.
Figure~\ref{fig:CDF_plot} shows the cumulative distribution of classifications as a function of confidence; this gives an indication of how the numbers change with a different confidence thresholds. 
We mainly use a fiducial $90\%$ confidence threshold for our quoted results.

\begin{table}[]
    \footnotesize
    \centering
        \begin{tabular}{l @{\extracolsep{3pt}}r r r r r r@{}} 
  \hline 
  & \multicolumn{3}{c}{Hanford} & \multicolumn{3}{c}{Livingston} \\ 
  \cline{2-4} \cline{5-7}
 \multicolumn{1}{c}{Gravity Spy class} &  \multicolumn{1}{c}{SNR $> 7.5$} & \multicolumn{1}{c}{$p > 90\%$} & \multicolumn{1}{c}{$p > 95\%$} & \multicolumn{1}{c}{SNR $> 7.5$} & \multicolumn{1}{c}{$p >90\%$} & \multicolumn{1}{c}{$p > 95\%$} \\ 
 \hline \hline 
1080 Lines          &   344 &    78 &    34 &   942 &   269 &   141 \\ 
1400 Ripples        &   253 &    85 &    49 &  7634 &  2384 &  1479 \\
Air Compressor      &   343 &   117 &    76 &  2901 &  1314 &   952 \\
Blip                &  7438 &  6020 &  5582 &  5554 &  4264 &  3873 \\
Blip Low Frequency  &  4042 &  2467 &  2059 & 21522 & 15614 & 14003 \\
Chirp               &    41 &     8 &     5 &    29 &    12 &     8 \\
Extremely Loud      & 13235 & 10938 & 10335 &  8994 &  7304 &  6835 \\
Fast Scattering     &  2243 &  1286 &  1118 & 74120 & 55211 & 50782 \\
Helix               &    91 &    15 &     9 &   229 &    37 &    16 \\
Koi Fish            & 11242 &  8447 &  7536 & 11153 &  7016 &  5800 \\
Light Modulation    &   146 &    45 &    29 &   753 &   191 &   133 \\
Low-frequency Burst & 21211 & 19410 & 18756 &  5771 &  3855 &  3448 \\
Low-frequency Lines &  3955 &  1536 &  1131 & 13749 &  3751 &  2125 \\
No Glitch           &  7783 &  5247 &  3874 & 14050 &  6748 &  4773 \\
Paired Doves        &   269 &    29 &    12 &  4079 &   277 &   130 \\
Power Line          &   303 &   164 &   135 &  1985 &  1441 &  1314 \\
Repeating Blips     &  1845 &  1078 &   902 &  1142 &   459 &   350 \\
Scattered Light     & 63333 & 57118 & 53701 & 57400 & 47258 & 43009 \\
Scratchy            &   643 &   367 &   311 &   444 &   287 &   263 \\ 
Tomte               &  1892 &  1360 &  1242 & 46144 & 39299 & 37573 \\
Wandering Line      &    30 &    10 &     5 &    64 &    28 &    20 \\
Whistle             &  6238 &  5371 &  5128 &  8623 &  6150 &  5721 \\ 
Violin Mode         &   884 &   436 &   366 &  1709 &   300 &   190 \\ 
 \hline 
\end{tabular}
    \caption{Number of Gravity Spy classifications in O3 \ac{LIGO} Hanford and Livingston data. 
    For each detector, the left column gives the total number of triggers with \ac{SNR} $> 7.5$ classified, regardless of the confidence of the classification, while the middle and right columns give the number of classifications with confidence $p > 90\%$ (our fiducial threshold) and $p > 95\%$, respectively.}
    \label{tab:O3_glitches} 
\end{table}

\begin{figure}
    \centering
    \includegraphics[width=6.0in]{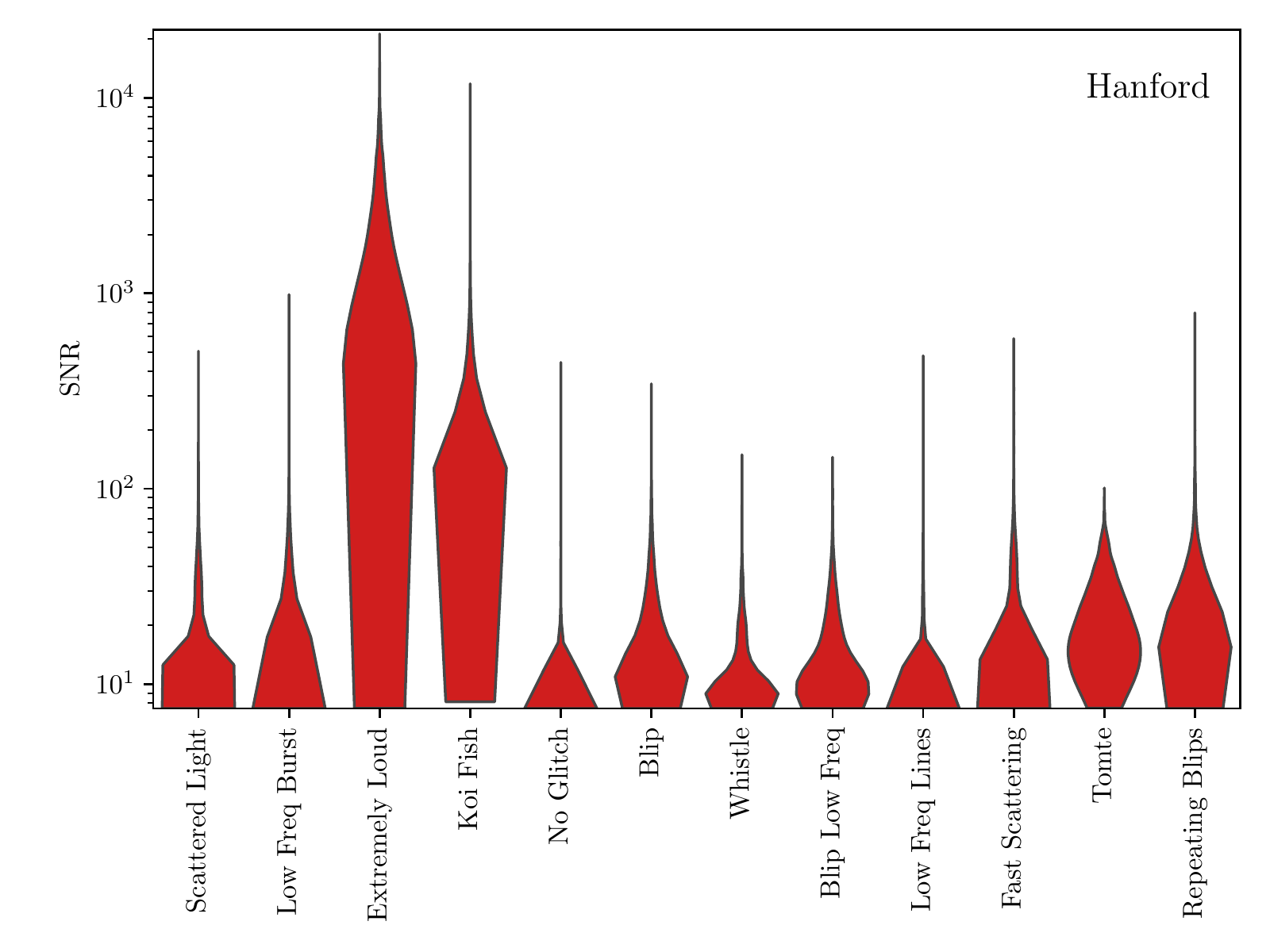}
    \caption{\ac{SNR} distributions for \ac{LIGO} Hanford glitches identified with a confidence $p > 90 \%$. 
    Only results for classes with a prevalence greater than $1 \%$ in Hanford data are shown. 
    The width of the distribution is normalized to be uniform across the different classes, and the classes are ordered in decreasing order of prevalence from left to right. 
    Table~\ref{tab:O3_glitches} lists the numbers of triggers in each class for the full list of classes, and analogous distributions for Livingston data are shown in Fig.~\ref{fig:L1_scatter}.}
    \label{fig:H1_scatter}
\end{figure}

\begin{figure}
    \centering
    \includegraphics[width=6.0in]{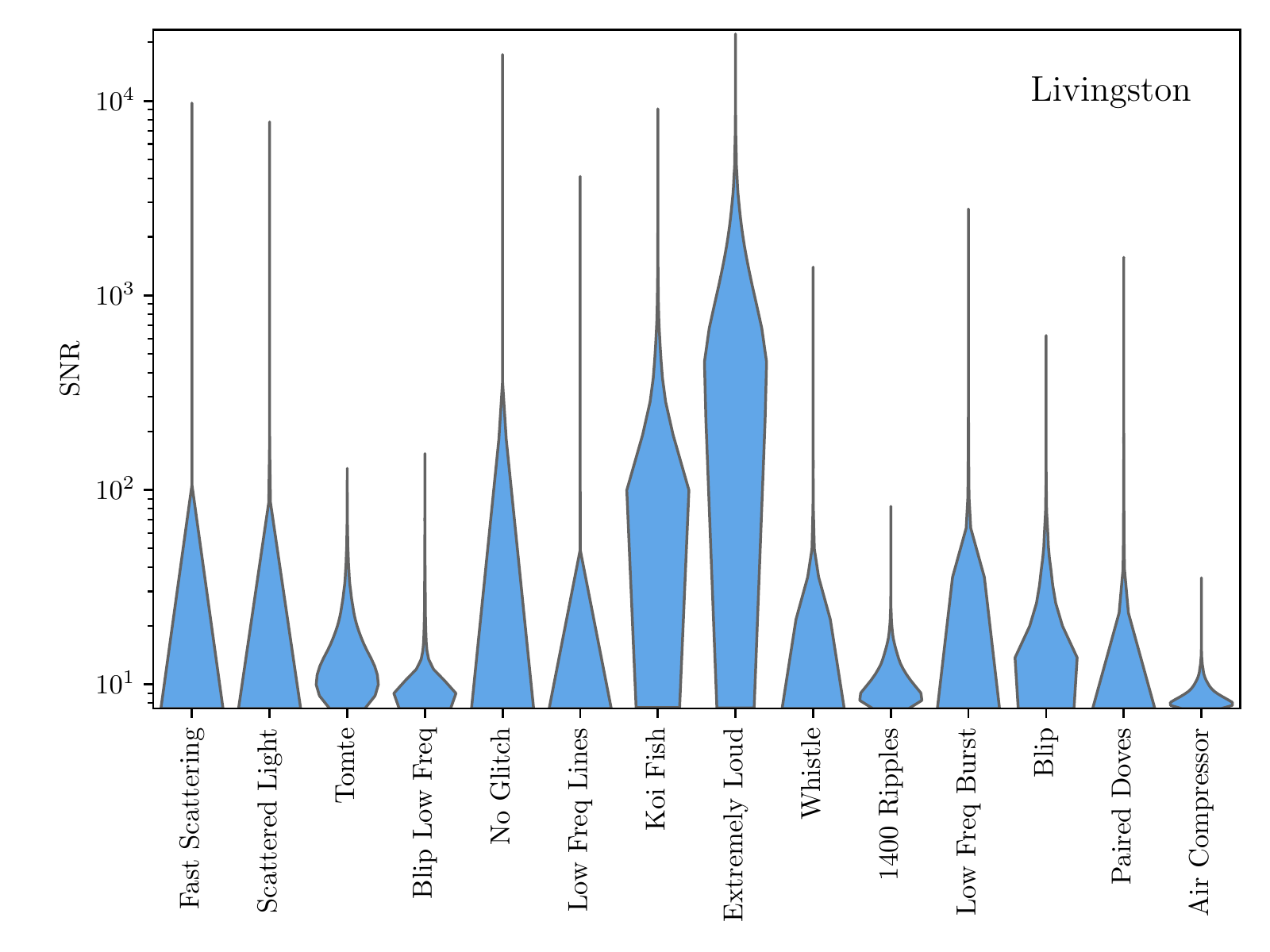}
    \caption{\ac{SNR} distributions for \ac{LIGO} Livingston glitches identified with a confidence $p > 90 \%$. 
    Only results for classes with a prevalence greater than $1 \%$ in Livingston data are shown. 
    The width of the distribution is normalized to be uniform across the different classes, and the classes are ordered in decreasing order of prevalence from left to right. 
    Table~\ref{tab:O3_glitches} lists the numbers of triggers in each class for the full list of classes, and analogous distributions for Hanford data are shown in Fig.~\ref{fig:L1_scatter}.}
    \label{fig:L1_scatter}
\end{figure}

The number of glitches and the split between classes differs between the two observatories. 
Figure~\ref{fig:H1_scatter} shows the O3 distribution of glitches as a function of \ac{SNR} for the most common classes (classes that have a $>1\%$ prevalence) in \ac{LIGO} Hanford data, and Fig.~\ref{fig:L1_scatter} shows the same for \ac{LIGO} Livingston.

During O3, the most common classes of glitches to occur at Livingston was due to scattered light~\cite{Accadia:2010zzb,Ottaway:2012oce,Valdes:2017xce}, specifically, Scattered Light (also known as Slow Scattering)~\cite{Soni:2020rbu} and Fast Scattering (also known as Crown)~\cite{Soni:2021cjy}. 
Approximately $27\%$ of all the glitches in O3 were classified as Fast Scattering by the Gravity Spy machine-learning analysis with a confidence of $> 90\%$. 
Scattered Light made up about $23\%$ of glitches with a Gravity Spy confidence of $> 90\%$. 
The relative motion between optical surfaces in \ac{LIGO} are strongly correlated with the presence of light scattering. 
The rate of Scattered Light glitches decreased during the second half of O3 (O3b) following the introduction of reaction-chain tracking in January 2020~\cite{LIGOScientific:2021djp}, which reduced the relative motion between the test-mass mirror and its counterpart used in control of the suspension system~\cite{Soni:2020rbu}.

Tomtes were another common glitch class for Livingston, making up approximately $19\%$ of all the glitches with a Gravity Spy confidence of $> 90\%$. 
The origins of these are currently unknown, as no environmental or instrumental couplings have been determined. They commonly appear with a frequency of $40~\mathrm{Hz}$, and repeat often over the course of one day~\cite{Davis:2021ecd}. 

At Hanford, Scattered Light, Low-frequency Bursts, and Extremely Loud glitches were the most common glitch classes. 
Reaction-chain tracking was also implemented at Hanford to help mitigate Scattered Light. 
Low-frequency Bursts were common during August 2019. 
Extremely Loud glitches are large disturbances to the detector and often cause big drops in the detector's astrophysical range (the distance out to which a source can be typically detected~\cite{Chen:2017wpg}). 
Scattered Light made up about $47\%$ of O3 glitches classified with $>90\%$ confidence at Hanford, while Extremely Loud and Low-frequency Bursts made up about $9\%$ and $16\%$, respectively.

\begin{figure}
    \centering
        \begin{subfigure}{0.9\textwidth}
            \centering
            \includegraphics[width= 1.0\textwidth]{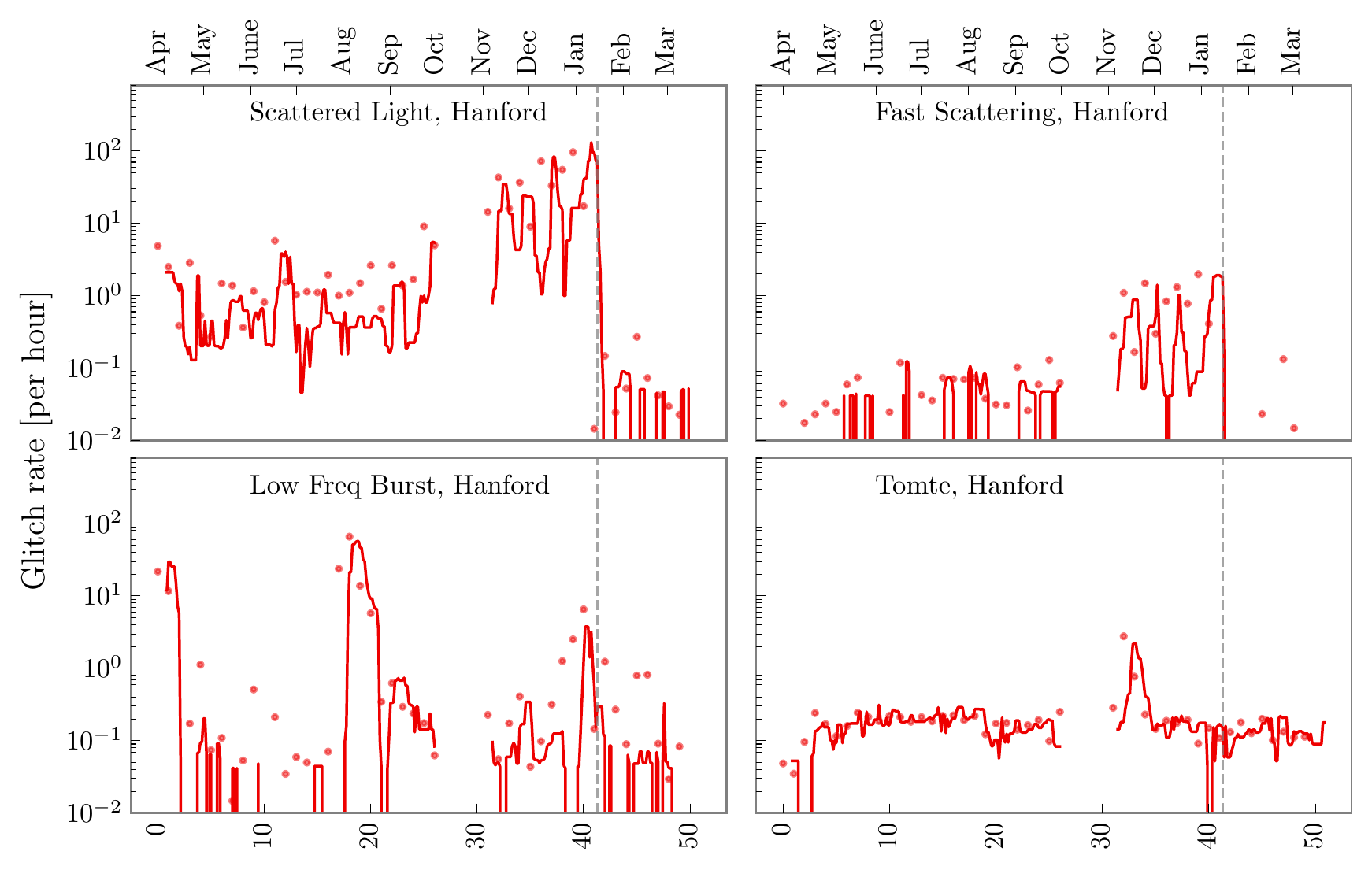}
        \end{subfigure}
        
        \begin{subfigure}{0.9\textwidth}
            \centering
            \includegraphics[width= 1.0 \textwidth]{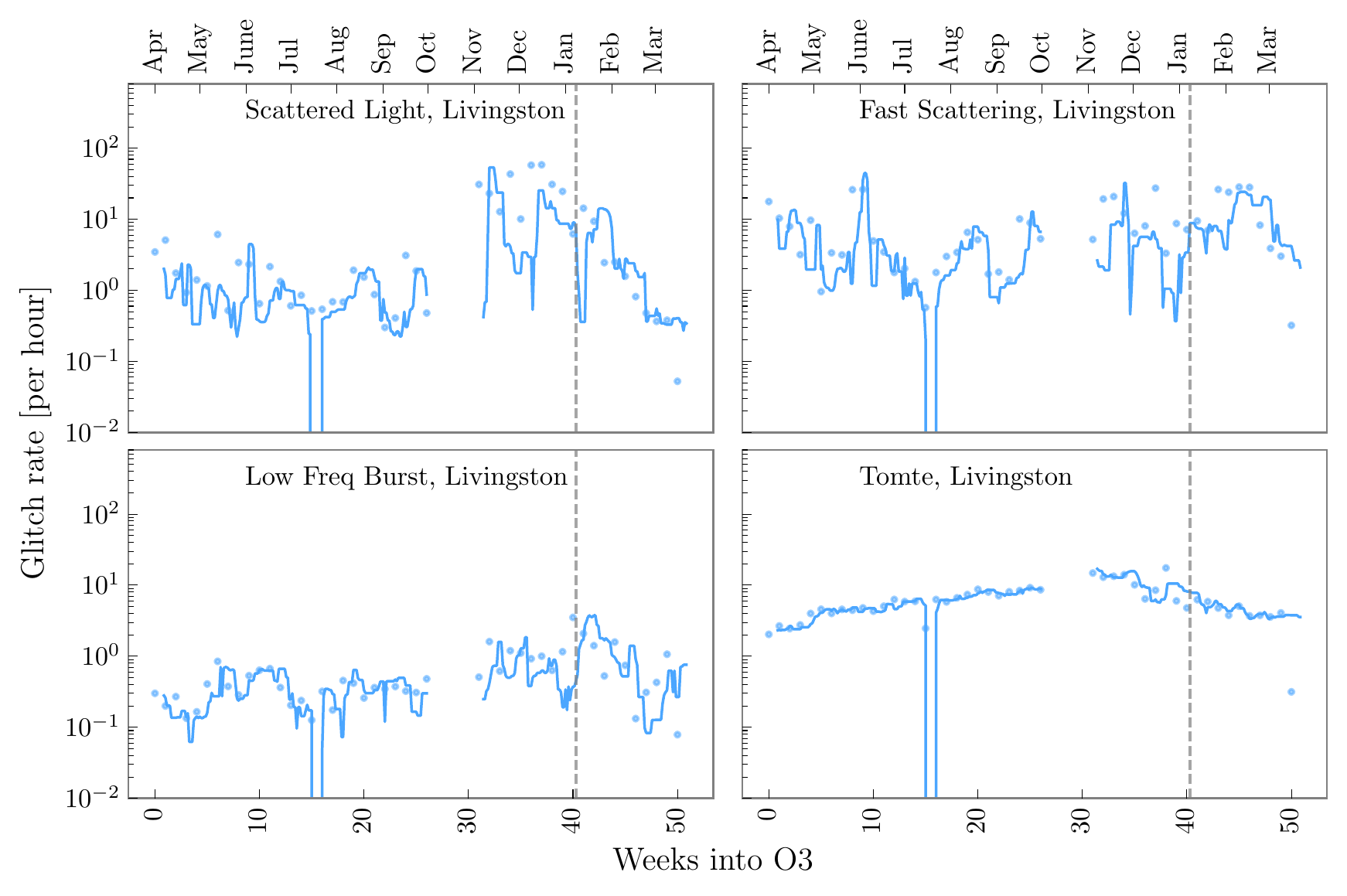}
        \end{subfigure}

    \caption{Hourly glitch rate (per unit observing time) for four glitch types (classified with confidence $> 90\%$) at \ac{LIGO} Hanford and \ac{LIGO} Livingston during O3 on different days of the week. 
    The rate is calculated as the number of glitches per unit observing time.
    The solid traces show the rolling median of the daily average glitch rate across seven day intervals, while the dots show the glitch rate for each calendar week. 
    The dashed vertical lines show the times when reaction-chain tracking was implemented~\cite{Soni:2020rbu,LIGOScientific:2021djp}. 
    The month of October was used for commissioning, and its data is not shown here.}
    \label{fig:weekly_glitch_rate}
\end{figure}

Figure~\ref{fig:weekly_glitch_rate} shows the hourly rate of four glitch classes (Scattered Light, Fast Scattering, Low-frequency Burst and Tomte) across the weeks of the O3 run for both Hanford and Livingston~\cite{Abbott:2020niy,LIGOScientific:2021djp}. 
The rate is calculated per unit observing time.
The glitch rates were calculated using those classified with $> 90\%$ confidence. 
This shows the large increase in Scattered Light glitches in the second part of the observing run and the subsequent reduction after the introduction of reaction-chain tracking~\cite{Soni:2020rbu,Davis:2021ecd,LIGOScientific:2021djp}. 

\begin{figure}
    \centering
        \begin{subfigure}{0.9\textwidth}
            \centering
            \includegraphics[width= 1.0\textwidth]{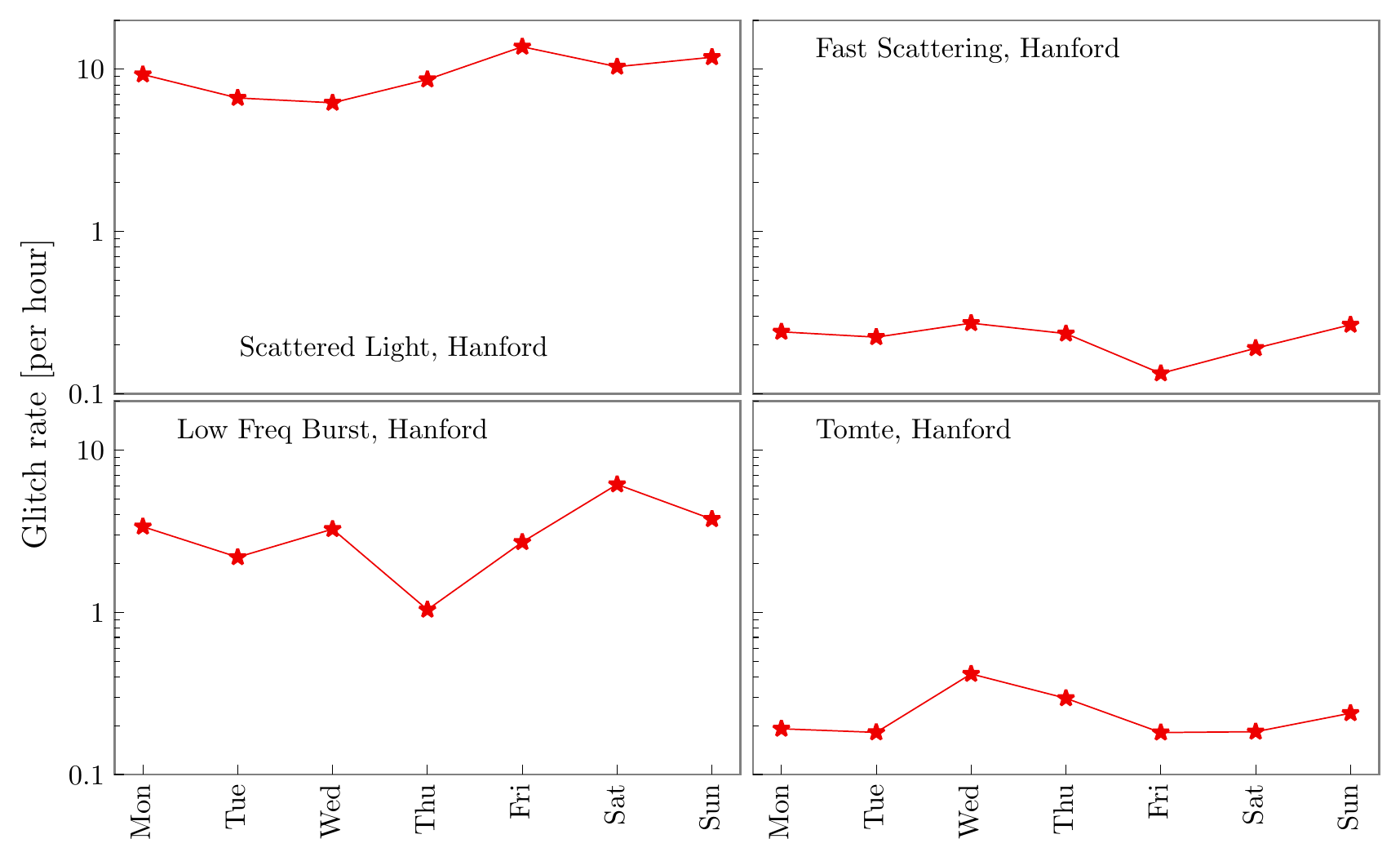}
        \end{subfigure}
        
        \begin{subfigure}{0.9\textwidth}
            \centering
            \includegraphics[width= 1.0\textwidth]{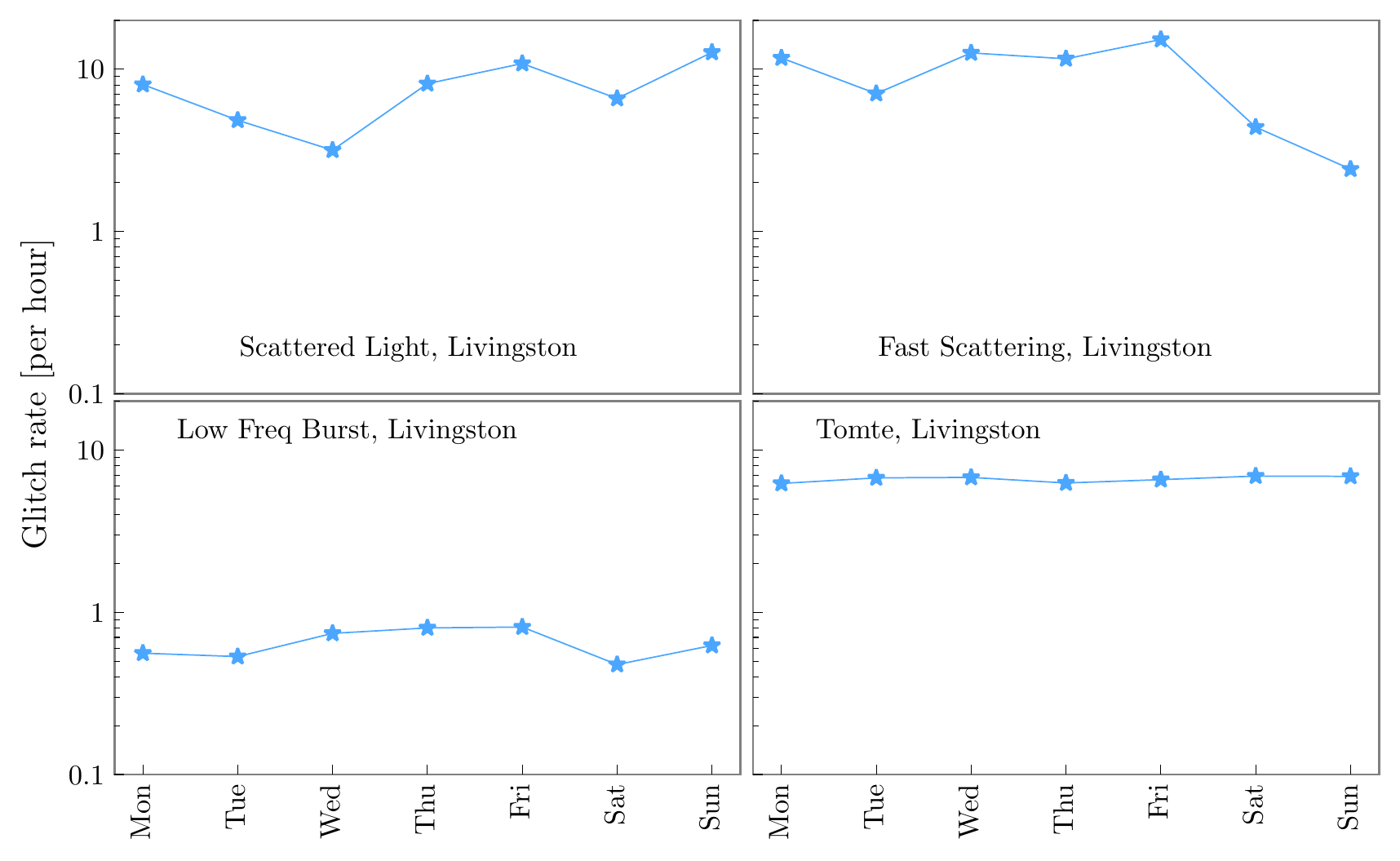}
        \end{subfigure}

    \caption{Hourly glitch rate for weekdays folded across the entire O3 run. 
    The rate is calculated as the number of glitches per unit observing time, and we plot the average over each weekday.
    The month of October was used for commissioning and its data is not shown here.  }
    \label{fig:folded_glitch_rate}
\end{figure}

Figure~\ref{fig:folded_glitch_rate} shows a different visualization of the variation in glitch prevalence with time: how the glitch rate (for the same classes shown in Fig.~\ref{fig:weekly_glitch_rate}) changes with the day of the week.%
\footnote{Plotting the number of glitches (the glitch rate multiplied by the detector duty cycle) instead of the glitch rate, would show a significant drop on Tuesdays, as this corresponds to the day of routine maintenance.}
Fast Scattering shows a decline during the weekend at \ac{LIGO} Livingston, as at these times there is less anthropogenic noise around the detectors. 
A similar difference is not visible at \ac{LIGO} Hanford because of the much lower rate of Fast Scattering transients at Hanford ($0.22$ per hour) compared to Livingston ($9.05$ per hour) during O3: a relatively higher ground motion in the anthropogenic band around Livingston makes Fast Scattering a much bigger problem there~\cite{Soni:2021cjy,LIGOScientific:2021djp}.
In contrast to Fast Scattering, Tomte shows negligible variation, indicating a lack of correlation with human activities.

\subsection{Data quality around candidates}\label{sec:candidates}

The data set includes glitch classifications for data around the time of several gravitational-wave candidates. 
This happens either when there is a glitch picked up by Omicron, if a gravitational-wave signal is loud enough to trigger Omicron, or if some combination of glitch and signal is identified. 
Here we review these Gravity Spy classifications, and illustrate both how Gravity Spy may identify glitches around candidates and how it may struggle in classifying a gravitational-wave signal.

Table~\ref{tab:O3a-events} and Table~\ref{tab:O3b-events} provide details of example candidates from the first and second parts of O3 (O3a and O3b), respectively, with associated Gravity Spy classifications. 
This list was compiled by cross-referencing the times associated with public alerts and high-significance candidates from offline analyses (whether or not they are identified as instrumental in origin)~\cite{Abbott:2020niy,LIGOScientific:2021usb,LIGOScientific:2021djp,LIGOScientific:2021tfm,LIGOScientific:2021hoh,KAGRA:2021bhs,Mishra:2022ott,Olsen:2022pin,Nitz:2021zwj} with the Gravity Spy data set. 
For this analysis, a time window of $\pm 5~\mathrm{s}$ around each candidate time was used to search for entries in the Gravity Spy data set.
The majority of candidates did not have a corresponding entry in the data set classified by Gravity Spy.

\begin{table}
    \small
    \footnotesize
        \begin{tabular}{l r l l} 
  \hline 
 \multicolumn{1}{c}{Superevent} & \multicolumn{1}{c}{Time} & \multicolumn{1}{c}{Gravity Spy classification}  & \multicolumn{1}{c}{Description} \\ 
 \hline\hline
S190930ak & 2019-09-30 23:46:50 & H: Scattered Light & {Instrumental origin}~\cite{LIGOScientific:2021djp} \\
& 2019-09-30 23:46:53 & H: Scattered Light &  \\
S190930s & 2019-09-30 13:35:37 & L: Low Frequency Lines & GW190930\_133541~\cite{GCN25870,Abbott:2020niy} \\
S190928c & 2019-09-28 02:11:45 & L: Tomte & Retracted~\cite{GCN25883,Abbott:2020niy} \\
S190924am & 2019-09-24 23:26:50 & L: Fast Scattering & Instrumental origin~\cite{LIGOScientific:2021tfm} \\ 
 & 2019-09-24 23:26:52 & L: Fast Scattering &  \\
 & 2019-09-24 23:26:54 & L: Fast Scattering &  \\
S190924h & 2019-09-24 02:18:42 & L: Tomte & GW190924\_021846~\cite{GCN25828,Abbott:2020niy} \\
S190910s & 2019-09-10 11:28:07 & L: Chirp & GW190910\_112807~\cite{Abbott:2020niy}  \\
S190904w & 2019-09-04 17:49:10 & L: Fast Scattering & Instrumental origin~\cite{Mishra:2022ott} \\ 
S190829u & 2019-08-29 21:05:56 & L: Koi Fish & Retracted~\cite{GCN25554,Abbott:2020niy} \\
S190814bv & 2019-08-14 21:10:38 & L: Scattered Light & GW190814\_211038~\cite{GCN25320,LIGOScientific:2020zkf,Abbott:2020niy} \\ 
S190808ae & 2019-08-08 22:21:21 & H: Low Frequency Burst & Retracted~\cite{GCN25295,Abbott:2020niy}  \\
S190804q & 2019-08-04 08:35:43 & L: Koi Fish & Instrumental origin~\cite{LIGOScientific:2021hoh,LIGOScientific:2021djp} \\
S190803e & 2019-08-03 02:26:59 & H: Low Frequency Burst & GW190803\_022701~\cite{Abbott:2020niy} \\
S190728q & 2019-07-28 06:45:12 & L: No Glitch & GW190728\_064510~\cite{GCN25183,Abbott:2020niy} \\
S190701ah & 2019-07-01 20:33:02 & L: Fast Scattering & GW190701\_203306~\cite{GCN24948,Abbott:2020niy} \\
S190630ag & 2019-06-30 18:52:05 & L: Chirp & GW190630\_18520~\cite{GCN24920,Abbott:2020niy} \\
{S190524q} & 2019-05-24 04:52:01 & L: No Glitch & {Retracted~\cite{GCN24655,Abbott:2020niy}} \\ 
 & 2019-05-24 04:52:02 & L: No Glitch &  \\
 & 2019-05-24 04:52:04 & L: No Glitch &  \\
 & 2019-05-24 04:52:09 & L: No Glitch & \\
S190521r & 2019-05-21 07:43:59 & H: Blip, L: Chirp & GW190521\_074359~\cite{GCN24629,Abbott:2020niy} \\
S190521g & 2019-05-21 03:02:29 & L: Blip Low Frequency & GW190521~\cite{GCN24618,LIGOScientific:2020iuh,Abbott:2020niy}  \\
S190519bj & 2019-05-19 15:35:44 & L: Blip & GW190519\_153544~\cite{GCN24598,Abbott:2020niy} \\
S190512at & 2019-05-12 18:07:18 & L: Tomte & GW190512\_180714~\cite{GCN24503,Abbott:2020niy} \\
S190430af & 2019-04-30 00:49:32 & H: Koi Fish & Instrumental origin~\cite{LIGOScientific:2021hoh} \\
S190421ar & 2019-04-21 21:38:53 & L: Power Line  & GW190421\_213856~\cite{GCN24141,Abbott:2020niy} \\
S190413ac & 2019-04-13 13:43:10 & L: Fast Scattering & GW190413\_134308~\cite{Abbott:2020niy} \\
S190412m & 2019-04-12 05:30:44 & L: Chirp & GW190412~\cite{GCN24098,LIGOScientific:2020stg,Abbott:2020niy}  \\
S190408an & 2019-04-08 18:18:06 & H: Low Frequency Burst & GW190408\_181802~\cite{GCN24063,Abbott:2020niy}  \\
 \hline 
 \end{tabular}
    \caption{Gravity Spy classifications coincident with confident, marginal and retracted O3a gravitational-wave candidates~\cite{Abbott:2020niy,LIGOScientific:2021usb,LIGOScientific:2021djp,LIGOScientific:2021tfm,LIGOScientific:2021hoh,KAGRA:2021bhs,Mishra:2022ott,Olsen:2022pin,Nitz:2021zwj}. 
    Equivalent results for O3b are shown in Table~\ref{tab:O3b-events}.
    The main Gravity Spy analysis uses data flagged by the Omicron pipeline as an input, and so only classifies a subset of candidates. 
    Omicron may pick up the candidate, a near-by glitch, or some combination of the two.
    The first column gives the corresponding candidate identification used in the Gravitational-wave Candidate Event Database (as used for low-latency alerts); the second gives the Coordinated Universal Time of the Omicron trigger ($\pm 5~\mathrm{s}$ from the time of the candidate); the third column gives the Gravity Spy classification with H and L indicating whether data from Hanford or Livingston, respectively, have been analysed; the fourth column gives details of the final status of the candidate (and citations).}
    \label{tab:O3a-events} 
\end{table}

\begin{table}
    \footnotesize
    \centering
        \begin{tabular}{l r l l} 
  \hline 
 \multicolumn{1}{c}{Superevent} & \multicolumn{1}{c}{Time} & \multicolumn{1}{c}{Gravity Spy classification}  & \multicolumn{1}{c}{Description} \\ 
 \hline\hline
S200311bg & 2020-03-11 11:58:53 & L: Blip & GW200311\_115853~\cite{GCN27358,LIGOScientific:2021djp} \\
S200224ca & 2020-02-24 22:22:34 & H: Blip, L: Chirp & GW200224\_222234~\cite{GCN27184,LIGOScientific:2021djp} \\
S200214br & 2020-02-14 22:45:26 & L: Fast Scattering & Instrumental origin~\cite{LIGOScientific:2021tfm,LIGOScientific:2021djp} \\ 
{S200129m} & 2020-01-29 06:55:00 & L: Fast Scattering & {GW200129\_065458~\cite{GCN26926,LIGOScientific:2021djp}} \\
 & 2020-01-29 06:54:58 & H + L: Chirp &  \\
S200121aa & 2020-01-21 03:17:48 & H: Blip & Instrumental origin~\cite{LIGOScientific:2021djp} \\ 
S200116ah & 2020-01-16 11:56:12 & L: Tomte & Retracted~\cite{GCN26785} \\
S200114f & 2020-01-14 02:08:18 & L: Tomte & Instrumental origin~\cite{GCN26734,LIGOScientific:2021hoh,LIGOScientific:2021tfm} \\
S200112r & 2020-01-12 15:58:38 & L: Chirp & GW200112\_155838~\cite{GCN26715,LIGOScientific:2021djp}  \\
S200108v & 2020-01-08 10:00:38 & L: Extremely Loud & Retracted~\cite{GCN26665} \\
S200106av & 2020-01-06 18:34:32 & H + L: Scattered Light & Retracted~\cite{GCN26641,LIGOScientific:2021djp} \\
S191225aq & 2019-12-25 21:57:15 & L: Tomte & Retracted~\cite{GCN26585,LIGOScientific:2021tfm} \\
S191223an & 2019-12-23 01:41:59 & L: Tomte & Instrumental origin~\cite{LIGOScientific:2021tfm} \\
S191213g & 2019-12-13 04:34:08 & L: Scattered Light & Unretracted, low significance~\cite{GCN26402,LIGOScientific:2021djp} \\
{S191212q} & 2019-12-12 08:27:25 & H: Scattered Light & {Retracted~\cite{GCN26394}} \\
 & 2019-12-12 08:27:28 & H: Scattered Light &  \\
{S191127p} & 2019-11-27 05:02:28 & H: Scattered Light & {GW191127\_050227~\cite{LIGOScientific:2021djp}} \\
 & 2019-11-27 05:02:24 & H: Scattered Light & \\
S191120aj & 2019-11-20 16:23:24 & L: Air Compressor & Retracted~\cite{GCN26263} \\
S191117j & 2019-11-17 06:08:22 & L: Extremely Loud & Retracted~\cite{GCN26254} \\
{S191113q} & 2019-11-13 07:17:53 & L: No Glitch & {GW191113\_071753~\cite{LIGOScientific:2021djp}} \\
 & 2019-11-13 07:17:48 & L: No Glitch &  \\
S191110x & 2019-11-10 18:08:42 & L: Koi Fish & Retracted~\cite{GCN26218} \\
{S191109d} & 2019-11-09 01:07:17 & H: Scattered Light, L: Blip & {GW191109\_010717~\cite{GCN26202,LIGOScientific:2021djp}} \\
 & 2019-11-09 01:07:15 & H: Scattered Light & \\
 & 2019-11-09 01:07:13 & L: Scattered Light &  \\
 & 2019-11-09 01:07:12 & H: Scattered Light &  \\
S191103a & 2019-11-03 01:25:52 & L: Tomte & GW191103\_012549~\cite{LIGOScientific:2021djp} \\
 \hline 
 \end{tabular}
    \caption{Gravity Spy classifications coincident with confident, marginal and retracted O3b gravitational-wave candidates~\cite{LIGOScientific:2021djp,LIGOScientific:2021tfm,LIGOScientific:2021hoh,KAGRA:2021bhs,Mishra:2022ott,Nitz:2021zwj}. 
    This is equivalent to Table~\ref{tab:O3a-events} but for O3b. 
    The first column gives the corresponding candidate identification used in the Gravitational-wave Candidate Event Database; the second gives the Coordinated Universal Time of the Omicron trigger ($\pm 5~\mathrm{s}$ from the time of the candidate); the third column gives the Gravity Spy classification with H and L indicating whether data from Hanford or Livingston, respectively, have been analysed; the fourth column gives details of the final status of the candidate (and citations).}
    \label{tab:O3b-events} 
\end{table}

\begin{figure}
    \centering
    \includegraphics[width=6.0in]{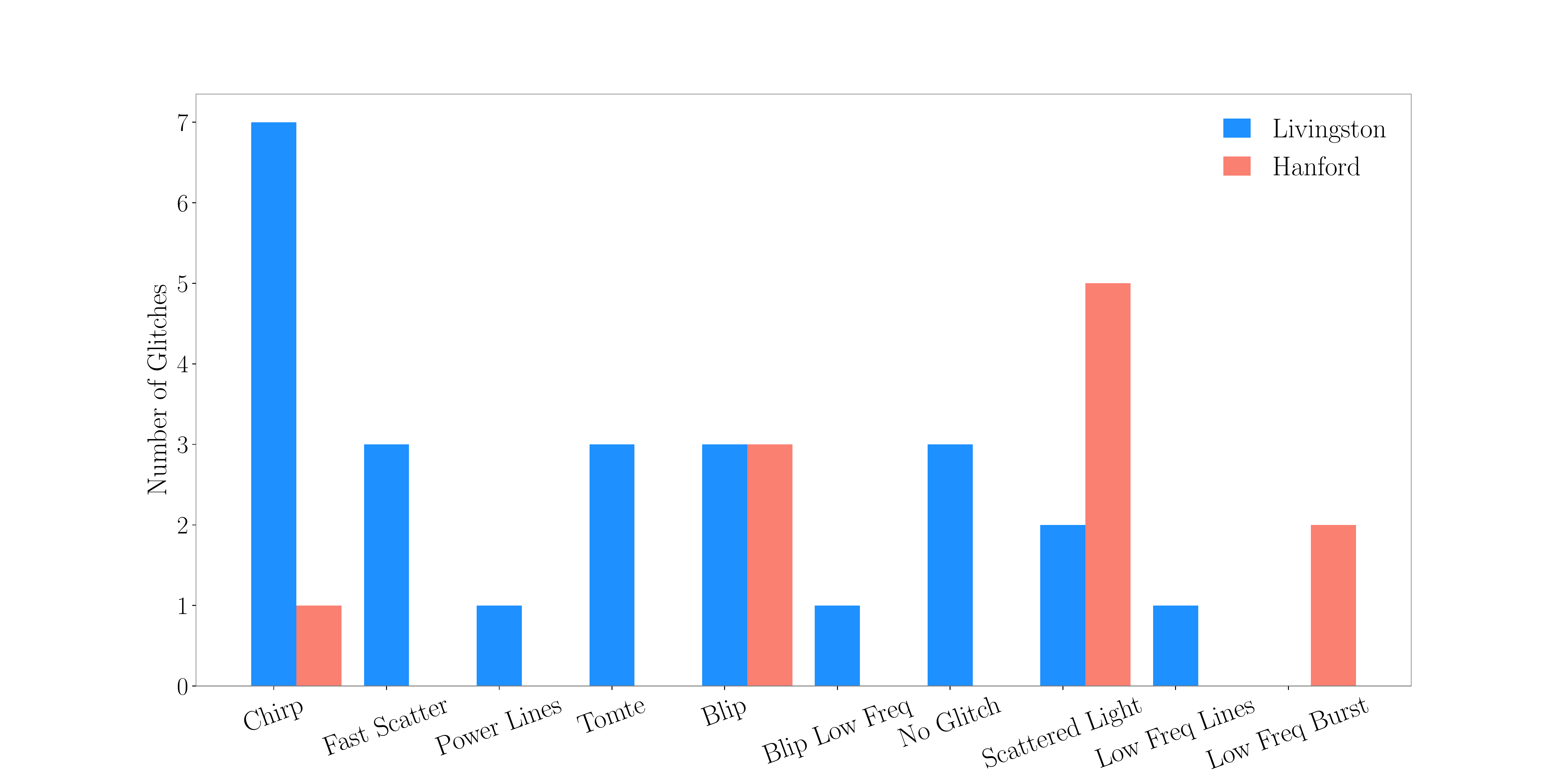}
    \caption{Gravity Spy classifications around O3 gravitational-wave candidates at \ac{LIGO} Hanford and Livingston. 
    For each candidate, a window of $\pm 5~\mathrm{s}$ used to identify entries in the Gravity Spy data set. 
    The machine-learning algorithm may be attempting to classify a gravitational-wave signal, a nearby glitch, or some combination of the two; it has not been trained to identify the full diversity of astrophysical gravitational-wave signals, nor how to classify data containing both a signal and a glitch.}
    \label{fig:glitchesH1L1}
\end{figure}

First, we consider the set of classifications around gravitational-wave candidates without an identified instrumental origin:
\begin{itemize}
\item From Livingston, there are $14$ O3a candidates that have at least one trigger identified by Gravity Spy, and $7$ O3b candidates. 
Three of the O3b events had two Livingston triggers during the time of the candidate. 
The most common class of glitches found were Chirps. 
Fast Scattering, Blip and Tomte were other common classifications. 

\item At Hanford, only $7$ candidates from O3 are part of the Gravity Spy data set. 
One of these candidates has three associated Hanford glitches, and another has two. 
The most common class to occur at times associated with these candidates was Scattered Light.

\item There were $4$ candidates in which a glitch was found at both detectors: GW190521\_074359, GW191109\_010717, GW200129\_065458 and GW200224\_222234. 
GW190521\_074359, GW200129\_065458 and GW200224\_222234 are amongst the highest \ac{SNR} candidates from O3~\cite{Abbott:2020niy,LIGOScientific:2021djp}. 
GW190521\_074359~\cite{Abbott:2020niy} and GW200224\_222234~\cite{LIGOScientific:2021djp} both have a Blip glitch identified at Hanford, and a Chirp at Livingston; while GW200129\_065458 has a Chirp at both, in addition to a Fast Scattering glitch at Livingston~\cite{LIGOScientific:2021djp}. 
For GW191109\_010717 there are Scattered Light glitches at both detectors, plus a Blip at Livingston~\cite{LIGOScientific:2021djp}.

\end{itemize}
The distribution of Gravity Spy classifications is shown in Fig.~\ref{fig:glitchesH1L1}.

\begin{figure}
    \centering
    \includegraphics[width=6.0in]{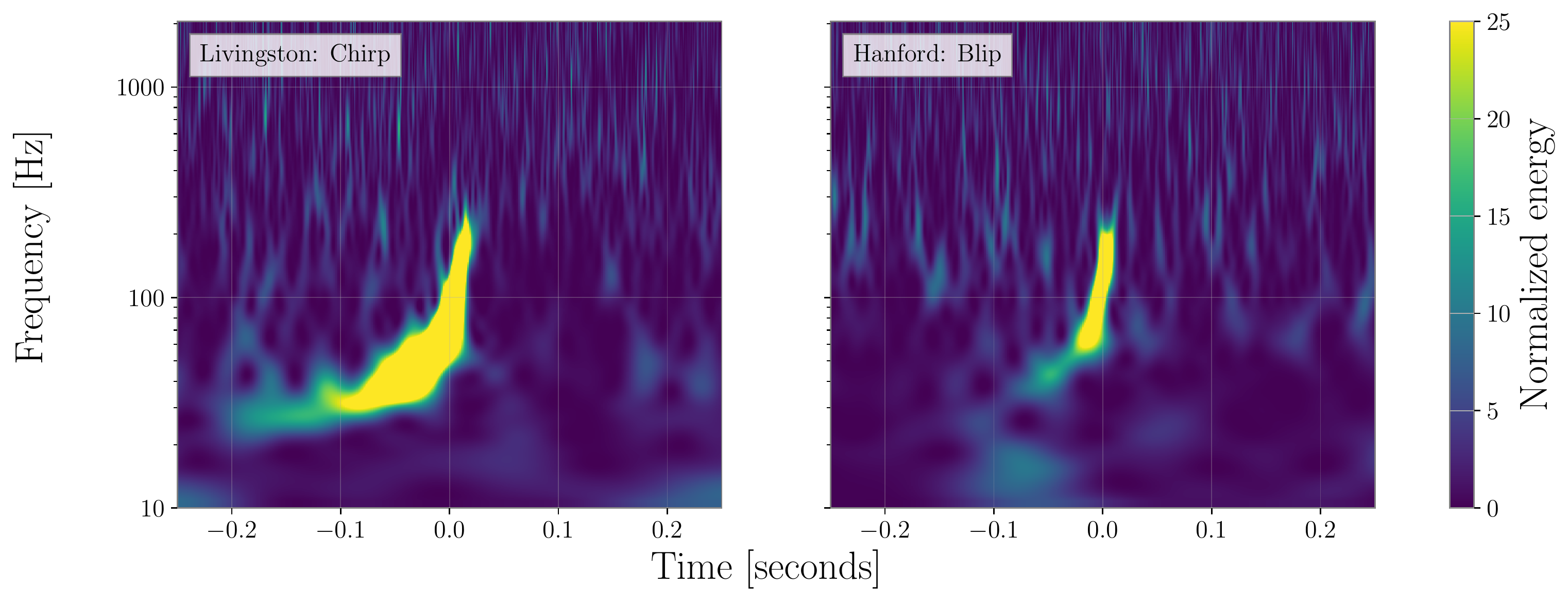}
    \caption{Gravitational-wave candidate GW190521\_074359~\cite{Abbott:2020niy}. 
    At Livingston, this glitch was classified as a Chirp, and at Hanford it was classified as a Blip. 
    The \ac{SNR} of the signal is higher in Livingston, which is why the chirp-like structure is easier to identify. }
    \label{fig:l1h1glitch}
\end{figure}

The Chirp class was originally created for hardware injections (simulated signals used for testing) representing compact binary coalescences~\cite{Biwer:2016oyg}, and hence might be expected to capture many of these candidates, as is the case. 
However, a chirp-like time--frequency morphology is only visible for the highest \ac{SNR} signals; as Livingston is the more sensitive detector, there are more high \ac{SNR} signals in its data. 
Tomte and Blip share a similar morphology to Chirps, and so may be confused for lower-\ac{SNR} signals. 
Figure~\ref{fig:l1h1glitch} illustrates an example (GW190521\_074359~\cite{Abbott:2020niy}) where a the higher-\ac{SNR} Livingston signal is classified as a Chirp, while the lower-\ac{SNR} Hanford signal is (mis)classified as a Blip. 

When a candidate is present at the same time as a glitch, it may be that the glitch is picked up by the classification algorithm. 
Data-quality checks~\cite{Davis:2022ird} indicated that data mitigation was needed for many candidates from O3 where there was excess noise overlapping the gravitational-wave signal.
GW190413\_134308, GW190701\_203306, GW190814 and GW200129\_065458 all required data mitigation for Livingston data, while GW191109\_010717 and GW191127\_050227 required data mitigation for Hanford data~\cite{Abbott:2020niy,LIGOScientific:2021djp}. 
These all correspond to cases where there is a Gravity Spy classification of a glitch outside of the Chirp--Blip--Tomte family in the relevant detector. 
However, there is not a perfect correlation between instances where data mitigation was required and Gravity Spy glitch classifications, and there are both candidates where mitigation was required, but there is no entry in the Gravity Spy data set, and candidates where there is a Gravity Spy glitch classification but no data mitigation was required. 
The former could happen if the excess noise was below the threshold for Omicron trigger, but still identified by the careful data-quality checks performed to evaluate candidates. 
The latter could happen if the noise is at a frequency that does not impact signal analysis (e.g., $\lesssim 20~\mathrm{Hz}$), or if the \ac{CNN} is confused by the combination of signal plus noise, and makes a misclassification. 
The Gravity Spy training set does not currently include examples of signals plus glitches.

To summarise, Gravity Spy is \emph{not} a detection algorithm, but a noise-classification algorithm. 
As such, it is not intended to discriminate between gravitational-wave signals and glitches. 
Most gravitational-wave signals are comparatively low in \ac{SNR}, making them more difficult to be picked up by Gravity Spy. 
Even when analysed by Gravity Spy, gravitational-wave signals will not all currently be put into the Chirp class. 
Consequently, the glitch classifications are contaminated (at a low rate) by gravitational-wave signals. 

Along with analyzing the O3 gravitational-wave candidates, we also looked at other candidates that were determined to be false alarms. 
During these events at Hanford, the most common glitch type seen was Scattered Light. 
At Livingston, there was more of a variety ranging from Tomtes, Koi Fish, Extremely Loud, Fast Scattering, and No Glitch. 

\begin{figure}
    \centering
    \includegraphics[width=4.5in]{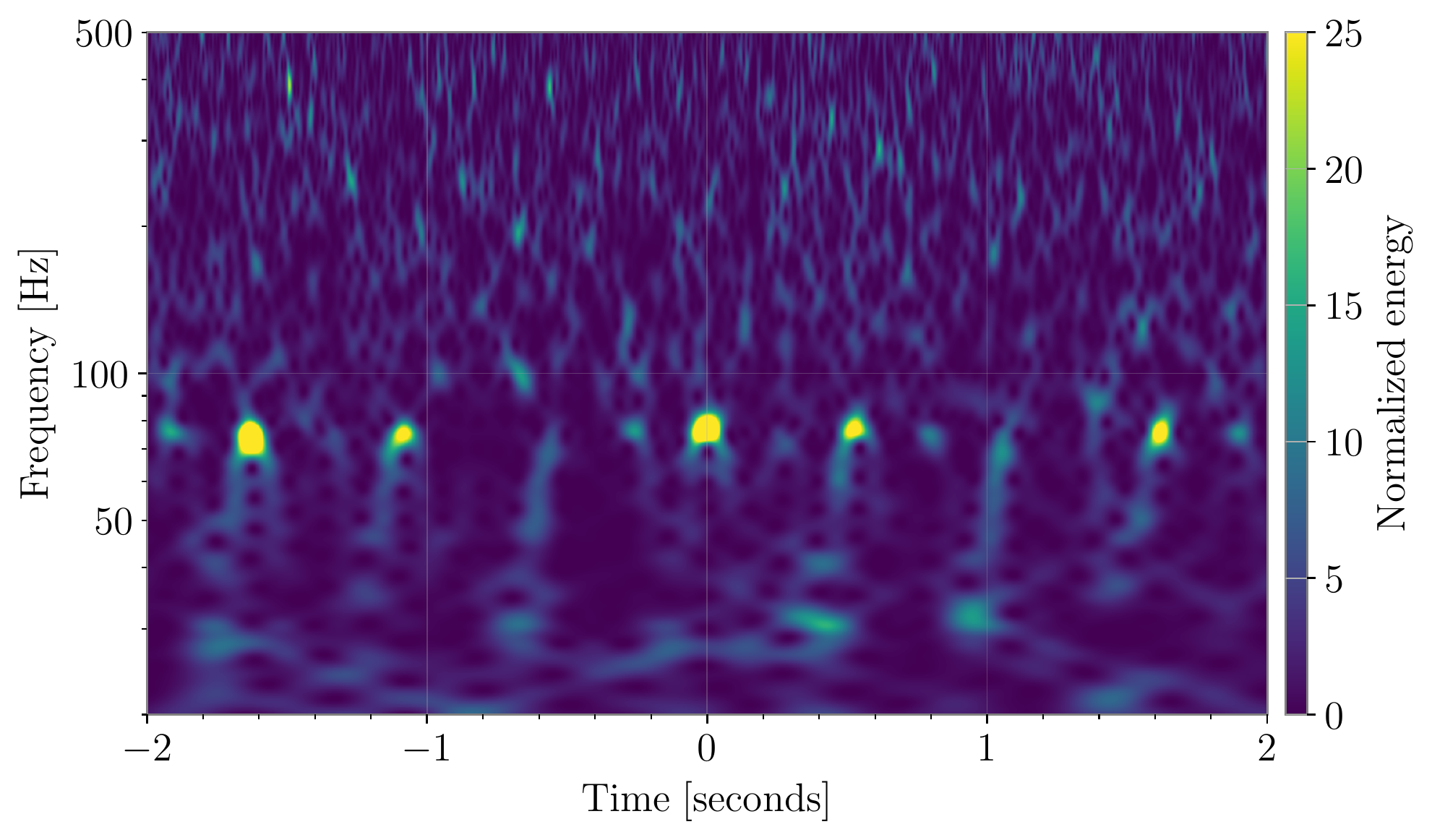}
    \caption{Example of a Livingston trigger classified as No Glitch from a time corresponding to the retracted candidate S190524q~\cite{GCN24655,Abbott:2020niy}. 
    Despite being labelled as No Glitch, the time--frequency resembles a high-frequency Fast Scattering glitch. 
    This trigger was classified by the Gravity Spy \ac{CNN} with a confidence of $94\%$.}
    \label{fig:noglitchexample}
\end{figure}

Of the candidates with an instrumental origin, the glitches classified as No Glitch are of particular interest: for the retracted candidate S190524q, there were 4 glitches classified as No Glitch. 
Figure~\ref{fig:noglitchexample} shows data around S190524q~\cite{GCN24655,Abbott:2020niy}, and despite the No Glitch classification, there is excess power visible. 
These glitches appear like a high-frequency analogue of Fast Scattering, which does not match any existing Gravity Spy class. 
This highlights how the existing set of classes does not catch the full diversity of noise in the detector, and that further refinements of the \ac{CNN} are needed to properly classify new types of glitches.

\subsection{Data release}\label{sec:data-release}

The data release of Gravity Spy machine-learning classifications is available from Zenodo~\cite{coughlin_scott_2021_5649212}. 
This consists of \ac{CSV} files for each detector and observing run (O1, O2, O3a and O3b). 
The \ac{CSV} files consist of columns describing: (i) metadata output from the Omicron pipeline~\cite{Robinet:2015om,Robinet:2020lbf} such as the time of the trigger, trigger peak frequency, bandwidth and amplitude, as well as the data analysed (the main gravitational-wave strain channel); (ii) the unique Gravity Spy identifier of the glitch; (iii) the machine-learning confidence for each of the original $22$ glitch categories; (iv) the machine-leaning classification and the confidence of this, and (v) links to Omega scans hosted by Zooniverse. 
Times are given as \ac{GPS} times, and can be used to identify the relevant data from the \ac{GWOSC}~\cite{LIGOScientific:2019lzm}.%
\footnote{\ac{GWOSC} \href{http://www.gw-openscience.org/}{gw-openscience.org/}} 
Examples of how to use the data release are given in a Python notebook accompanying the release.

\section{Discussion}\label{sec:discussion}

The \ac{LIGO} detectors in Livingston, Louisiana and Hanford, Washington nominally share an identical design~\cite{TheLIGOScientific:2014jea}, and thus we might not expect their performance to differ much from each other. 
However, due to differences in their commissioning progress~\cite{Abbott:2020qfu,LIGOScientific:2017bnn,Buikema:2020dlj}, and in their surrounding environments, the two observatories do differ in practice~\cite{Abbott:2016xvh,LIGOScientific:2018mvr,Davis:2021ecd,Abbott:2020niy,LIGOScientific:2021djp}.
For example, due to the presence of extra low-frequency noise at Hanford during O3, its sensitivity is about a factor of $2$ lower in the frequency band $20$--$60~\mathrm{Hz}$, as compared to Livingston~\cite{Buikema:2020dlj}.  
Additionally, the amount of ground motion in the anthropogenic ($1$--$6~\mathrm{Hz}$) and microseism ($0.1$--$0.5~\mathrm{Hz}$)  bands is usually larger near Livingston than near Hanford. 
Consequently, there can be considerable difference in the amount and nature of transient noise between the two detectors: during O3b, the rate of Omicron transients with \ac{SNR} above $10$ at Livingston was about $1.7$ times higher than at Hanford. 

\begin{figure}[h]
\captionsetup[subfigure]{}
   \centering
    \begin{subfigure}{0.65\textwidth}
        \centering
         \includegraphics[width= 1.1\textwidth]{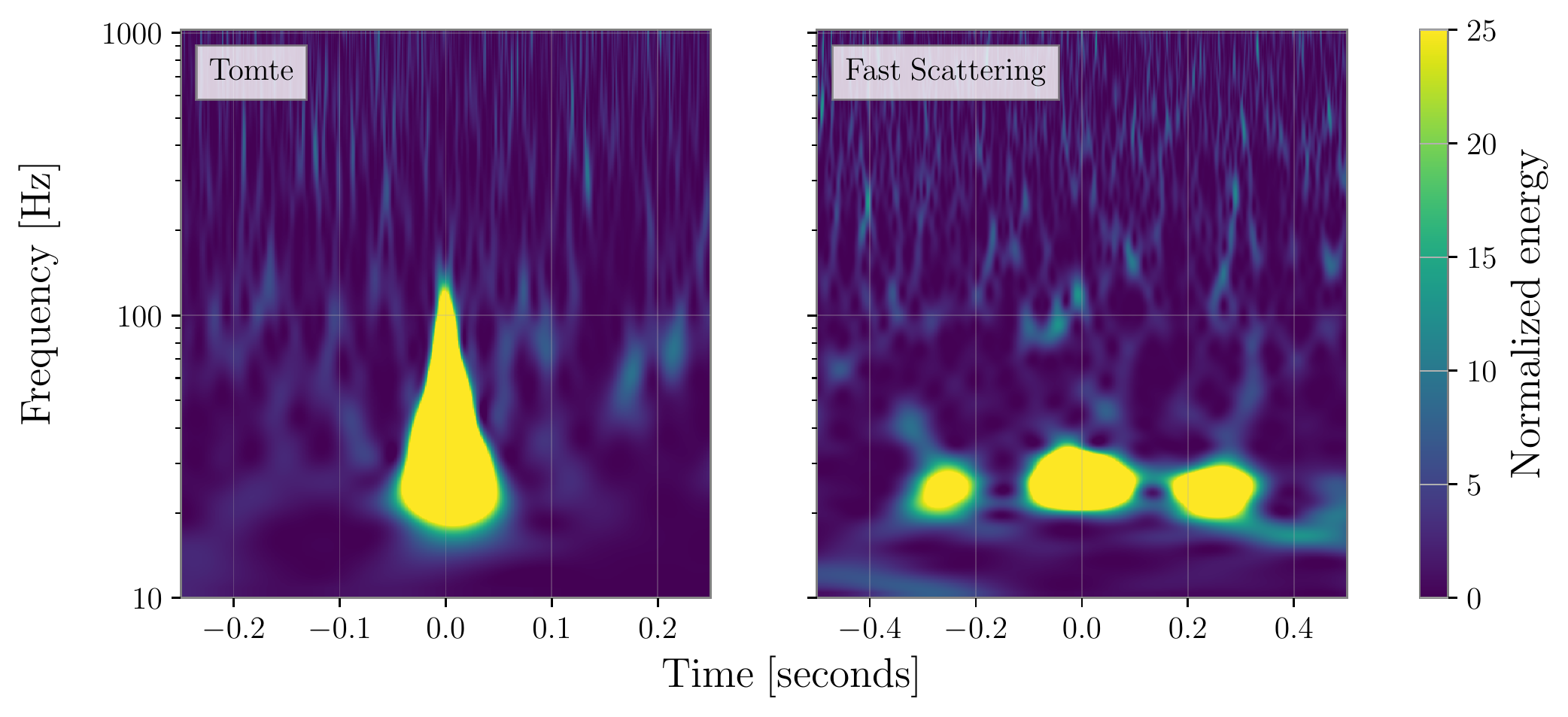}
         \label{fig:tomteFS_scan}
    \end{subfigure}
    \begin{subfigure}{0.60\textwidth}
        \centering
         \includegraphics[width =1.1\textwidth]{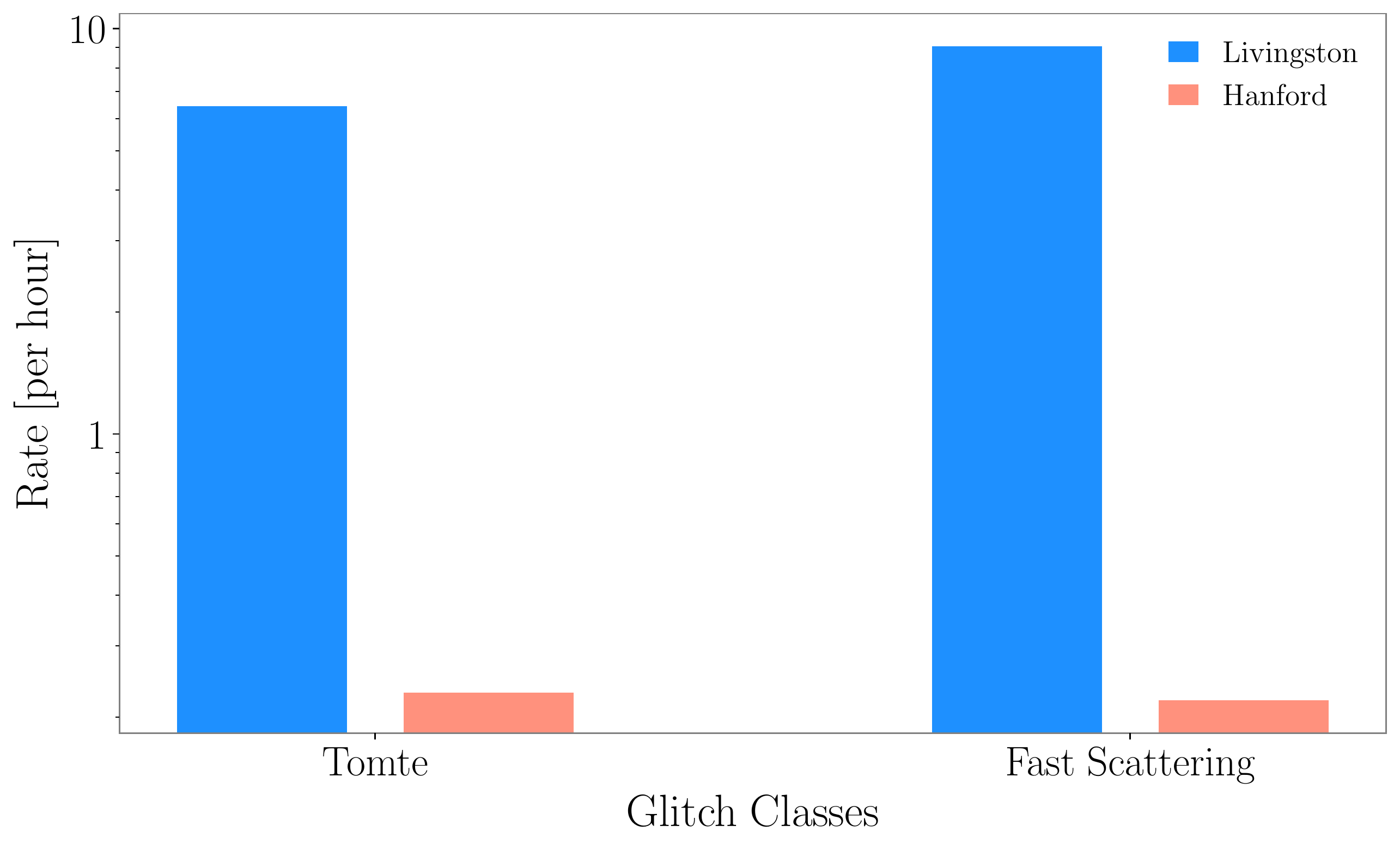}
         \label{fig:tomteFS_comparison}
    \end{subfigure}
    \caption{Time--frequency morphology of the glitch categories Tomte and Fast Scattering shown in the \emph{top} plot. 
    Both of these classes were more common at Livingston than at Hanford during O3, as shown in plot on the \emph{bottom}. }
    \label{fig:Tomte_FS}
\end{figure}
We see a difference in the number and distribution of glitches across the different Gravity Spy classes (e.g., Table~\ref{tab:O3_glitches}).
For example, during O3, the glitch classes Tomte and Fast Scattering were more common in Livingston, and this increased prevalence boosted the overall glitch rate~\cite{alog:Tomte_L1,Davis:2021ecd,Soni:2021cjy}. 
Examples of these two glitches classes, and a comparison of their prevalence during O3 is shown in Fig.~\ref{fig:Tomte_FS}.

Fast Scattering was first noticed as a significant source of noise during the engineering runs preceding O3~\cite{alog:44803,Soni:2021cjy}. 
The prevalence of Fast Scattering was a primary motivation for updating the Gravity Spy model to include new classes for the analysis of O3 data~\cite{Soni:2021cjy}. 
Nearly all Fast Scattering during O3 is below $\sim60~\mathrm{Hz}$. 
This transient noise is linked to an increase in ground motion in the anthropogenic and microseism bands near the detector~\cite{dcc:noise-sprint,FS_Corner_Coupling}. 
These two bands are usually noisier at Livingston than at Hanford, and this (combined with the differences in the detectors' low-frequency sensitivity) meant that Fast Scattering was more common at Livingston ($9.05$ per hour) than at Hanford ($0.22$ per hour)~\cite{alog:ashley,Davis:2021ecd}.

Unlike Fast Scattering, we have not yet been able to identify an environmental or instrumental coupling that can explain the origin of Tomte glitches. 
There are ongoing detector characterisation efforts to understand how this glitch may couple in the detector~\cite{alog:Tomte_L1}. 
While we do not know the origins of Tomte glitches, we do observe a difference in their prevalence at the two observatories: during O3, the rate of Tomte glitches at Livingston was $6.44$ per hour, while at Hanford the rate was $0.23$ per hour. 
Tomte glitches have most of their power below $\sim64~\mathrm{Hz}$. 
The difference in the low-frequency sensitivity between the two detectors may be partially responsible for the difference in the rates during O3. 
Further study of when Tomte glitches occur, and the differences between Livingston and Hanford, may reveal the origins of these glitches.

A successful example of detector characterisation during O3 was the identification of the source of Scattered Light (Slow Scattering) glitches, and its subsequent mitigation~\cite{Soni:2020rbu}.
Scattered Light glitches have a significant impact on data quality because they occupy a large region time--frequency parameter space. 
As shown in Fig.~\ref{fig:glitches_combined}, Scattered Light transients appear as long-duration arches in spectrograms. 
These arches are characteristic of noise caused by light scattering. 
While the frequency gives some information on the motion of the component scattering the light, it is still difficult to identify the troublesome light path in the detectors.  
The Gravity Spy analysis played a significant role in understanding the source of Scattered Light: the occurrence of glitches classified as Scattered Light was found to correlate with motion of the the quad suspension~\cite{Davis:2021ecd,Soni:2020rbu}, which is captured by the optical shadow sensors and magnetic actuators (OSEMS)~\cite{Aston:2012ona,Matichard:2015eva}, indicating that the source of light scattering was part of the suspension system. 
The motion was subsequently reduced by employing reaction-chain tracking, which resulted in a considerable reduction in the rate of Scattered Light for the same degree of ground motion near the observatories~\cite{Soni:2020rbu}.
The resulting drop in the glitch rate is visible in Fig.~\ref{fig:weekly_glitch_rate}. 
This decline in the glitch rate of Scattered Light is sharper at Hanford than at Livingston due comparatively higher ground motion near Livingston in the microseism band during February 2020~\cite{Davis:2021ecd,LIGOScientific:2021djp}.

The fourth observing run (O4) will see the use of new and improved technologies~\cite{Cahillane:2022pqm}. 
Among them are frequency-dependent squeezing, new Faraday isolators, new test mass mirrors at Livingston, and higher laser power. 
These improvements will translate to a higher instrument sensitivity, thereby increasing our astrophysical reach for detecting gravitational-wave signals. 
However, a more sensitive detector is not just more sensitive to gravitational waves, it is also more sensitive to environmental and instrumental noise artifacts.
Compared to O2, the rate of glitches during O3a was four times higher at Livingston~\cite{Abbott:2020niy}. 
Like O3, it is possible that in O4 we will witness one or more new types of noise transients, and that these will appear only at one of the detectors. 
This could require using a \emph{site-specific} Gravity Spy training set and \ac{CNN} model to properly characterise O4 data quality.
The current plan for O4 is to sample the transients for any new glitch morphologies during the engineering run preceding O4, and retrain Gravity Spy before observing starts.

\section{Summary}\label{sec:summary}

Understanding data quality is a key aspect of gravitational-wave detector characterisation. 
The Gravity Spy machine-learning algorithm enables automated classification of segments of \ac{LIGO} data suspected to contain transient noise. 
Gravity Spy is routinely used in studies of data quality~\cite{Davis:2021ecd}, has been integral in the identification of new classes of glitches~\cite{Soni:2021cjy}, and has aided in the identification of the sources of glitches~\cite{Soni:2020rbu}.
Here we describe the data release of classifications for O1, O2 and O3.
Using \ac{CNN} models trained for O1--O2~\cite{Zevin:2016qwy,Bahaadini:2018git} and for O3~\cite{Soni:2021cjy}, we have analysed Advanced LIGO data from these first three observing runs; the results are publicly available from Zenodo~\cite{coughlin_scott_2021_5649212}. 
These can be used for a range of studies, from investigating environmental and instrumental origins of glitches, to developing new data-analysis pipelines; we have used the Gravity Spy classifications to illustrate some of the properties of data quality in O3 (as well as highlighting some limitations of the data set).

This release covers data from O1--O3. 
O4 (and subsequent observing runs)~\cite{Abbott:2020qfu} will follow improvements to the detector that may lead to the appearance of new glitch classes (and possibly the elimination of current glitch classes). 
Therefore, the Gravity Spy machine-learning model may need to be updated to account for these changes. 
To aid detector-characterisation experts in identifying new glitch classes and building a training set of example glitches, we will draw upon the Zooniverse volunteers along with machine-learning clustering approaches. 
Gravity Spy volunteers have previously rapidly identified new classes based upon their time--frequency morphologies~\cite{Soni:2021cjy}, and for O4 we will support their investigations into the causes of glitches by providing them with additional auxiliary channel data. 
Following the update of glitches classes, we anticipate that the classifications provided by the Gravity Spy project will enable further studies of LIGO data quality and improvements to data-analysis pipelines.

\ack{}

We thank the citizen-science volunteers of Gravity Spy who have contributed to the classifications of LIGO data. 
We are grateful to Marissa Walker and the anonymous referees for comments on the manuscript. 
Gravity Spy is partly supported by the National Science Foundation (NSF) award INSPIRE 1547880 and partially by award IIS-2107334. 
This work is supported by the NSF under grant PHY-1912648. 
JG is supported by NSF grant PHY-2110509. 
SB acknowledges support by NSF grants PHY-1912648 and IIS-2107334. 
SS acknowledges support of the NSF grant PHY-1764464 to the LIGO Laboratory.
MZ is supported by NASA through the NASA Hubble Fellowship grant HST-HF2-51474.001-A awarded by the Space Telescope Science Institute, which is operated by the Association of Universities for Research in Astronomy, Incorporated, under NASA contract NAS5-26555. 
CPLB acknowledges support from the CIERA Board of Visitors Research Professorship, and Science and Technology Facilities Council (STFC) grant ST/V005634/1. 
OP is supported by NSF grant PHY-1559694. 
VK was partially supported through a CIFAR Senior Fellowship, NSF grant PHY-1912648, and by Northwestern University.
This material is based upon work supported by NSF's LIGO Laboratory which is a major facility fully funded by the National Science Foundation.
This research has made use of data and software obtained from \ac{GWOSC} (\href{http://www.gw-openscience.org/}{gw-openscience.org}), a service of LIGO Laboratory, the LIGO Scientific Collaboration, the Virgo Collaboration, and KAGRA. 
The authors gratefully acknowledge the support of the United States NSF for the construction and operation of the LIGO Laboratory and Advanced LIGO as well as STFC of the United Kingdom, and the Max-Planck-Society for support of the construction of Advanced LIGO. 
Additional support for Advanced LIGO was provided by the Australian Research Council. 
Advanced LIGO was built under award PHY-0823459.
LIGO was constructed by the California Institute of Technology and Massachusetts Institute of Technology with funding from the National Science Foundation and operates under Cooperative Agreement PHY-1764464. 
This work used computing resources at CIERA funded by NSF grant PHY-1726951, and the computational resources and staff contributions provided for the Quest high performance
computing facility at Northwestern University which is jointly supported by the Office of the Provost, the Office for Research, and Northwestern University Information
Technology.
The authors are grateful for computational resources provided by the LIGO Laboratory and supported by National Science Foundation Grants PHY-0757058 and PHY-0823459. 
This document has been assigned LIGO document number \href{https://dcc.ligo.org/LIGO-P2200238/public}{LIGO-P2200238}. 
The data that support the findings of this study are openly available from Zenodo~\cite{coughlin_scott_2021_5649212}.

\appendix

\section{Glitch classes}\label{ap:classes}

\begin{figure}
    \centering
    \includegraphics[width=1\textwidth]{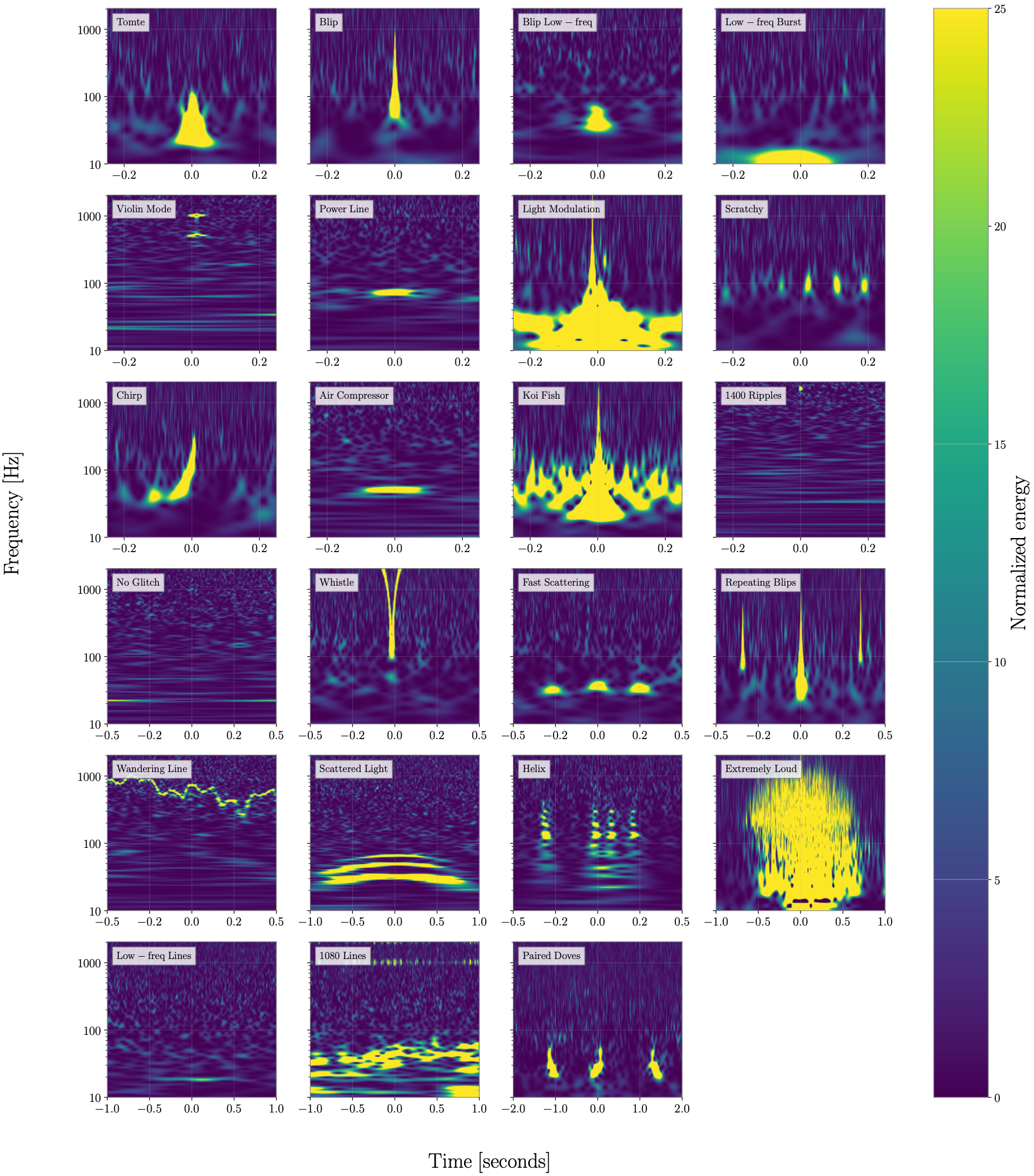}
    \caption{Time--frequency morphology for examples of the Gravity Spy classes in O3. 
    The classes are grouped by the time duration ($0.5~\mathrm{s}$, $1~\mathrm{s}$, $2~\mathrm{s}$ or $4~\mathrm{s}$) that best illustrates their features.
    \emph{First row}:  {Tomte},  {Blip},  {Blip Low Frequency} and  {Low-frequency Burst} ($0.5~\mathrm{s}$). 
    \emph{Second row}:  {Violin Mode},  {Power Line},  {Light Modulation} and  {Scratchy} ($0.5~\mathrm{s}$). 
    \emph{Third row}:  {Chirp},  {Air Compressor},  {Koi Fish} and  {1400 Ripples} ($0.5~\mathrm{s}$). 
    \emph{Fourth row}:  {No Glitch},  {Whistle},  {Fast Scattering} and  {Repeating Blips} ($1~\mathrm{s}$). 
    \emph{Fifth row}:  {Wandering Line},  {Scattered Light},  {Helix} ($1~\mathrm{s}$) and  {Extremely Loud} ($2~\mathrm{s}$). 
    \emph{Sixth row}:  {Low-frequency Lines},  {1080 Lines} and  {Paired Doves} ($4~\mathrm{s}$).  
    The Blip Low Frequency and Fast Scattering classes are not used for O1 and O2, but the O1 and O2 results do include an additional None of the Above class.}
    \label{fig:allglitch}
\end{figure}

The Gravity Spy projects classifies images into a range of classes. 
For \ac{LIGO} data from O1 and O2, $22$ classes are used in the \ac{CNN} model~\cite{Zevin:2016qwy,Bahaadini:2018git}, and for data from O3 $23$ classes (the older classes except None of the Above, plus Fast Scattering and Blip Low Frequency) are used~\cite{Soni:2021cjy}. 
In alphabetical order, the set of classes are,:
\begin{enumerate}
    \item \emph{1080 Lines}: These appear as short-duration dots repeating every $\sim0.1~\mathrm{s}$ at $\sim 1080~\mathrm{Hz}$. 
    They are also accompanied by noise below $64~\mathrm{Hz}$.
    These glitches were prevalent in Hanford date early in O2, but were reduced following improvements in the output mode cleaner~\cite{alog:1080-OMC}. 
    
    \item \emph{1400 Ripples}: These glitches appear as short ($\lesssim0.05~\mathrm{s}$) wavy lines at $\sim 1400~\mathrm{Hz}$. 
    
    \item \emph{Air Compressor}: This class appears as thick flat line at $\sim50~\mathrm{Hz}$. 
    In Hanford, these were found to be related to air compressor motors at the end stations~\cite{alog:21436}, and were reduced following the replacement the vibration isolators.
    
    \item \emph{Blip}: Blip glitches are broadband with very short ($\sim0.04~\mathrm{s}$) duration. 
    Due to their teardrop morphology, Blips can adversely influence the search for high-mass binary black hole signals. 
    Despite being a common glitch class, the cause of Blips is currently unknown~\cite{Cabero:2019orq}.
    
    \item \emph{Blip Low Frequency}: Otherwise known as \emph{Low-frequency Blips}, these glitches have a similar morphology to Blip glitches, except they occur at lower frequencies with peak frequencies $\sim10$--$50~\mathrm{Hz}$~\cite{Soni:2021cjy}. 
    This is a new glitch class added for O3.
    
    \item \emph{Chirp}: The characteristic sweep from low frequencies to high of a coalescing compact-object binary. 
    The class originally contained examples of simulated signals created by hardware injections~\cite{Biwer:2016oyg}. 
    The Chirp training set was created early in the era of gravitational-wave astronomy to accommodate hardware injections, and is not representative of our current understanding of the population of coalescing binaries~\cite{LIGOScientific:2021djp,LIGOScientific:2021psn}.
    
    \item \emph{Extremely Loud}: These broadband transients are characterised by very high \ac{SNR}, often leading to the spectrograms appearing saturated. 
    These correspond to large disturbances to the detectors, and may often be accompanied by a drop in the astrophysical range of the detector. 
    High-\ac{SNR} glitches from other classes (e.g., Koi Fish) may be classified as Extremely Loud.
    
    \item \emph{Fast Scattering}: Otherwise known as \emph{Crown}, these glitches appear as short-duration ($\sim0.2$--$0.3~\mathrm{s}$) arches~\cite{Soni:2021cjy}. 
    These arches often appear in groups, each separated by either $0.25~\mathrm{s}$ or $0.5~\mathrm{s}$. 
    They are correlated with ground motion in the anthropogenic ($1$--$6~\mathrm{Hz}$) band, which is usually caused by bad weather or human activity.
    This is a new glitch class added for O3, and they were the most common glitch in Livingston data.
    
    \item \emph{Helix}: These are broadband glitches, usually in the frequency region $16$--$512~\mathrm{Hz}$, often occurring in groups of two or three glitches separated from each other by $\sim0.1~\mathrm{s}$. 
    They may be related to glitches in the auxiliary lasers used to calibrate the detectors~\cite{alog:21436}. 
    
    \item \emph{Koi Fish}: These glitches are high-\ac{SNR} broadband glitches. 
    They typically occupy the frequency band $\sim20$--$1000~\mathrm{Hz}$, and can resemble Blips, but with pectoral fins at $\sim30~\mathrm{Hz}$. 
    
    \item \emph{Light Modulation}: These transients are usually high \ac{SNR}, with most of the noise content at $16$--$128~\mathrm{Hz}$, but there may also be one or more broadband spikes. 
    They are caused by amplitude fluctuations in the control signal of the optical sidebands used to regulate the length and alignment of optical cavities~\cite{TheLIGOScientific:2016zmo}.
    
    \item \emph{Low-frequency Burst}: These are usually short-duration ($\sim 0.25~\mathrm{s}$) transients between $\sim10$--$20~\mathrm{Hz}$, often appearing as a hump at the bottom of the spectrogram. 
    They were common at Livingston data during O1 and Hanford data in O3a. 
    
    \item \emph{Low-frequency Lines}: These appear mostly as flat lines, extending $\sim1.5$--$2~ \mathrm{s}$ in time and usually below  $\sim20~\mathrm{Hz}$. 
    
    \item \emph{No Glitch}: This category is used for Omicron triggers where there is not visible excess power in the  Gravity Spy spectrogram. 
    These are usually low-\ac{SNR} Omicron triggers, but can include short-duration, high-frequency ($\gtrsim 2000~\mathrm{Hz}$) transients than are difficult to resolve because of the logarithmic frequency scale used for the spectrograms.
    
    \item \emph{None of the Above}: This category is a catch-all for glitches that do not fit into the other categories. 
    Accordingly, there is no typical morphology. 
    This class is primarily useful when Zooniverse volunteers are classifying images. 
    This class was \emph{not} used for the final \ac{CNN} classification of O3 data.
    
    \item \emph{Paired Doves}: These appear as a pair of short duration transients, alternating between increasing and decreasing in frequency, with a separation of $\sim0.1~\mathrm{s}$. 
    These glitches are potentially related to periods of excess motion of the beamsplitter~\cite{alog:27138}.
    
    \item \emph{Power Line}: These glitches appear as narrow, flat lines, usually $\sim0.2$--$0.5~\mathrm{s}$ close to $60~\mathrm{Hz}$ (or harmonics of this). 
    This frequency corresponds to the electric power-grid frequency in United States, and glitches can be caused by a range of equipment that runs of this power supply~\cite{alog:23483,alog:32389}.
    
    \item \emph{Repeating Blips}: This class consists of multiple Blip-like glitches, often repeating with a cadence of $\sim0.25$--$0.50~\mathrm{s}$.
    
    \item \emph{Scattered Light}: Otherwise known as \emph{Slow Scattering} (to distinguish from Fast Scattering), they appears as long-duration ($\sim2.0$--$2.5~\mathrm{s}$) arches in the spectrograms.
    They are correlated with ground motion in the earthquake ($0.03$--$0.1~\mathrm{Hz}$) or microseism ($0.1$--$0.5~\mathrm{Hz}$) frequency bands. 
    In O3, it was found that Scattered Light was caused by the relative motion between the optical suspension system's end test-mass chain and the reaction-mass chain~\cite{Soni:2020rbu}.
    
    \item \emph{Scratchy}: Sometimes known as \emph{Blue Mountains}, these appear as a series of sharp peaks at intermediate frequencies $\sim60$--$250~\mathrm{Hz}$. 
    There may be $\sim10$--$30$ peaks per second. 
    They are related to light scattering from the Swiss cheese baffles~\cite{alog:36147,alog:43177}.
    
    \item \emph{Tomte}: These are short-duration glitches with a characteristic triangular shape. 
    They are similar to Blip or Blip Low-frequency glitches, and typically occupy the frequency band $\sim16$--$150~\mathrm{Hz}$. 
    They can adversely influence the search for high-mass binary black hole signals.
    
    \item \emph{Violin Mode}: These appear as disturbances at $\sim500~\mathrm{Hz}$ and harmonics. 
    These frequencies correspond to the resonances of the glass fibres that are used to suspend the mirrors.
    
    \item \emph{Wandering Line}: These long-duration transients have an undulating line morphology. 
    They can cover a wide range of frequencies, with multiple lines appearing at once at different frequencies, but are usually above $\sim256~\mathrm{Hz}$. 
    
    \item \emph{Whistle}: Also known as \emph{Radio Frequency Beat Notes}, these appear as U-, V- or W-shaped transients, typically above $\sim128~\mathrm{Hz}$ with most of the noise content above $\sim500~\mathrm{Hz}$. 
    They are caused when radio-frequency signals beat with the voltage controlled oscillators~\cite{Nuttall:2015dqa}.

\end{enumerate}
Examples for the $23$ classes used for O3 classification are shown in Figure~\ref{fig:allglitch}. 

In addition to the classes used in the \ac{CNN}, there are additional \ac{LIGO} glitch classes that have been proposed by Zooniverse volunteers during O3 that have not yet been incorporated into the machine-learning framework:
\begin{enumerate}
    \item \emph{70 Hz Line}: These appear as lines similar to Air Compressor or Power Line glitches, but centred at $\sim70~\mathrm{Hz}$.
    
    \item \emph{High-frequency Burst}: These appear as very short-duration transients at frequencies $\gtrsim1000~\mathrm{Hz}$. 
    
    \item \emph{Pizzicato}: These appear as a short ($\sim0.05~\mathrm{s}$) transient that resembles a flying saucer centered around $\sim500~\mathrm{Hz}$, $\sim1000~\mathrm{Hz}$, or both. 
    The frequencies correspond to violin modes of the suspension fibres, and the glitch may be related violin mode damping mechanisms, but the exact cause is yet to be identified.
\end{enumerate}
These, and further classes, may be added to the \ac{CNN} for future studies.

\bibliographystyle{iopart-num}
\bibliography{Gravity Spy.bib}

\end{document}